\documentstyle[12pt]{article}

\newcommand{\be}{\begin{equation}} 
\newcommand{\ee}{\end{equation}} 
\newcommand{\bea}{\begin{eqnarray}} 
\newcommand{\eea}{\end{eqnarray}} 
\newcommand{\nn}{\nonumber} 
\newcommand{\p}[1]{(\ref{#1})} 
 
\def\theequation{\arabic{section}.\arabic{equation}} 
\begin{document} 
\begin{titlepage} 
\begin{flushright}
LNF-00/009 (P)\\
JINR-E2-99-341 \\
hep-th/0003154 \\
15 March 2000
\end{flushright} 
\vskip 0.6truecm 
\begin{center}{\Large\bf 
$N=(4,4)$, $2D$ supergravity in $SU(2)\times SU(2)$ harmonic superspace}
\end{center} 
 \vskip 0.6truecm 
\centerline{\bf S. Bellucci${}^{\, a,1}$, E. Ivanov${}^{\,b,c,2}$} 
\vskip 0.6truecm 

\centerline{$^a${\it INFN, Laboratori Nazionali di Frascati, P.O. Box 
13, I-00044 Frascati, Italy}}

\vspace{0.1cm}
\centerline{$^b$ {\it
Laboratoire de Physique Th\'eorique et des Hautes Energies,}}
\centerline{\it Unit\'e associ\'ee au CNRS UMR 7589,~Universit\'e Paris 7}
\centerline{\it 2 Place Jussieu, 75251 Paris Cedex 05, France}

\vspace{0.1cm}
\centerline{$^c${\it Bogoliubov Laboratory of 
Theoretical Physics, JINR,}} 
\centerline{\it 141 980 Dubna, Moscow region, 
Russian Federation} 
\vskip 0.6truecm  \nopagebreak

\begin{abstract} 
\noindent We work out the basics of conformal $N=(4,4)$, $2D$ supergravity in 
the $N=(4,4)$, $2D$ analytic harmonic superspace with two independent 
sets of harmonic variables. We define the relevant most general 
analytic superspace diffeomorphism group and show that in the flat 
limit it goes over into the ``large'' $N=(4,4)$, $2D$ superconformal group. 
The basic objects of the supergravity considered are analytic vielbeins 
covariantizing two analyticity-preserving harmonic derivatives. For 
self-consistency they should be constrained in a certain way. We 
solve the constraints and show that the remaining irreducible field 
content in a WZ gauge amounts to a new short $N=(4,4)$ Weyl supermultiplet. 
As in the previously known cases, it involves no auxiliary fields 
and the number of remaining components in it coincides with the number 
of residual gauge invariances. We discuss various truncations of this 
``master'' conformal supergravity group and its compensations via couplings to 
$N=(4,4)$ superconformal matter multiplets. Besides recovering the standard 
minimal off-shell $N=(4,4)$ conformal and Poincar\'e supergravity 
multiplets, we find, at the linearized level, several 
new off-shell gauge representations. 

\end{abstract} 
\vfill
{\it E-Mail:}\\
{\it 1) bellucci@lnf.infn.it}\\
{\it 2) eivanov@lpthe.jussieu.fr,  eivanov@thsun1.jinr.ru}
 
\newpage 

\end{titlepage}

\section{Introduction} 
For building up self-consistent string models with 
$N=(4,4)$ worldsheet supersymmetry (SUSY) it is of primary importance 
to explore in full the structure of the relevant worldsheet 
conformal supergravity (SG), both on and off shell, as well as its couplings 
to $N=(4,4)$, $2D$ superconformal sigma models. In components 
and in the standard $N=(4,4)$, $2D$ superspace these issues were 
addressed in refs. [1]-[10].
Recently, there was a revival  of interest
to $N=(4,4)$ superconformal $2D$ theories caused by the  fact that they
describe the low-energy limits of some string theory  compactifications (see.
e.g., [11]-[13]).  
This makes it urgent to revert to the
problem of finding out an adequate  superspace description of $N=(4,4)$ SG and
listing all possible versions  of the latter.  

Here we present the basics of conformal $N=(4,4)$, $2D$  SG
in the analytic harmonic $SU(2)\times SU(2)$ superspace \cite{ISu}
with two independent sets of harmonic variables (for the left and 
right light-cone sectors). This kind of harmonic superspace is indispensable
for the off-shell description of $N=(4,4)$ supersymmetric torsionful sigma
models, with all supersymmetries being manifest. Our construction in its
starting points  follows the analogous one for conformal SG in  the analytic
subspace of $N=2$, $4D$ harmonic superspace [15]-[17],  
but eventually we find a few essential
differences from the latter  theory. These differences amount to a number of
novel features of our  construction compared to the existing approaches to
$N=(4,4)$, $2D$ SG.

First, in the $SU(2)\times SU(2)$ harmonic superspace {\it three} 
different SG 
groups containing local $SU(2)_L\times SU(2)_R$ symmetry can be defined 
($L$ and $R$ stand for the left- and right-handed 
$2D$ light-cone sectors). Two of them have as the rigid 
limits two different infinite-dimensional $N=4$, $SU(2)$ superconformal 
groups \cite{Ademollo} the realization of which in the flat 
harmonic superspace was given in \cite{{ISu}}. 
A closure of these two rigid superconformal groups is 
the large $N=4$, $2D$ superconformal group 
with the $SO(4)\times U(1)$ affine Kac-Moody group as internal 
symmetry (in each of two $2D$ light-cone sectors) [19]-[22]. 
The most general SG group  
which can be defined in the analytic $SU(2)\times SU(2)$ 
harmonic superspace yields 
in the flat limit just this large $N=(4,4)$ superconformal group. The 
corresponding SG can be treated as a ``master theory'' producing two 
$N=(4,4)$, $SU(2)$ SG theories as its proper truncations. 
Another, more elegant way of getting $N=(4,4)$, $SU(2)$ SG theories from the 
master $N=(4,4)$ SG is to couple the latter to appropriate 
harmonic superfield compensators. We explicitly demonstrate how 
one of $N=(4,4)$, $SU(2)$ SG groups can be recovered using this 
compensation procedure. The relevant compensator is one of 
the $SU(2)\times SU(2)$ 
harmonic superfields defined in \cite{IS3} (it contains $(32 + 32)$ off-shell 
components and generalizes the so-called nonlinear 
supermultiplet \cite{holl}). It should 
be stressed that the most characteristic 
feature of the master $N=(4,4)$ SG group is the presence of 
{\it four} local 
$SU(2)$ symmetries (corresponding to gauging left and right $SO(4)$) and 
two local $U(1)$ symmetries (corresponding to gauging left and 
right $U(1)$). The versions of off-shell conformal $N=(4,4)$ SG
known until now contained at most two local $SU(2)$ symmetries and no 
local $U(1)$ symmetries at all.
 
One more difference from the $N=2$, $4D$ case 
stems from the presence of two independent sets 
of $SU(2)$ harmonic variables in the $SU(2)\times SU(2)$ harmonic superspace. 
This peculiarity gives rise, on the one hand, to the 
property that the relevant groups of analytic superdiffeomorphisms 
are more powerful than their $N=2$, $4D$ counterpart, in the sense 
that they allow 
to gauge away more fields from the 
basic geometric objects of the theory, analytic vielbeins which 
covariantize two analyticity-preserving harmonic derivatives. 
On the other hand, 
prior to any gauge fixing, we are 
led to impose the constraints on the analytic vielbeins reflecting the 
commutativity of two independent analyticity-preserving harmonic 
derivatives in the flat case. The constraints and the original 
SG gauge group together work in such a way that in the WZ gauge 
we are left with no auxiliary fields at all. Besides, 
the number of gauge fields coincides with that of the 
remaining 
independent gauge parameters in the left and right sectors. Thus,  
the analytic vielbeins in the considered case actually describe a sum of 
two pure gauge Weyl multiplets. 
This sum can be naturally called the $N=(4,4)$ Beltrami-Weyl (BW) 
multiplet (the $SU(2)$ or $SO(4)\times U(1)$ one, depending on from  
which 
superdiffeomorphism group one proceeds). For the $N=(4,4)$, $SU(2)$ case 
our results agree with those of Schoutens \cite{schout}, who constructed 
the corresponding SG in the component approach by directly gauging 
the product of left and right $N=4$, $SU(2)$ superconformal groups. 
The standard conformal $N=(4,4)$ SG group corresponds to gauging 
the maximal finite-dimensional subgroup $SU(1,1|2)\times SU(1,1|2)$ 
of this product \cite{{PVN},{PPVN}}, \cite{dWPVN}-\cite{Gates2} 
and also gives rise to Weyl multiplet 
containing no off-shell degrees of freedom. A novel point is 
that this phenomenon 
of the one-to-one correspondence between 
the gauge fields and residual gauge invariances is continued as well 
to the more general case of $N=(4,4)$, $SO(4)\times U(1)$ SG group.
The supermultiplet of what is usually referred to as ``the minimal 
off-shell Poincar\'e $N=4$ SG'' \cite{{Gates1},{Gates2}} arises after 
coupling $N=4$, $SU(2)$
BW multiplet to a compensating 
superfield which represents one of twisted chiral 
multiplets in the analytic harmonic $SU(2)\times SU(2)$
superspace.  Thus the minimal $N=(4,4)$ SG 
representation corresponds to the two successive compensations: firstly, the 
$N=(4,4)$, $SO(4)\times U(1)$ SG group is compensated down to its 
$N=(4,4)$, $SU(2)$ subgroup by using some special harmonic 
compensator and, secondly, this subgroup is further compensated down 
to the group 
corresponding to the minimal representation via coupling to 
a twisted $N=(4,4)$ supermultiplet.  The 
existence of a dual formulation of the twisted multiplet with an 
infinite number of auxiliary fields \cite{ISu}  
implies the existence of new off-shell version of $N=(4,4)$ 
Poincar\'e SG  with an {\it infinite} number of auxiliary 
fields. 

In the present paper we do not aim to present the whole formalism 
of $N=(4,4)$ SG in harmonic superspace. We concentrate on describing the 
analytic superspace geometry 
of the $SU(2)$ and $SO(4)\times U(1)$, $N=(4,4)$ BW supermultiplets: 
define the relevant groups, the 
analyticity-preserving harmonic derivatives and the covariant 
constraints on the latter, and show that after choosing 
appropriate WZ gauges and solving the constraints we are left 
with the needed irreducible field contents. We present the invariant 
couplings of $SO(4)\times U(1)$, $N=(4,4)$ 
BW multiplet to the compensating $N=(4,4)$ multiplets, such that 
the residual gauge freedom is just one of the $N=(4,4)$, $SU(2)$ SG 
groups. Then we extend this coupling to include an arbitrary 
number of self-interacting twisted multiplets. 
We also show, at the linearized level, how to extract another 
$N=(4,4)$, $SU(2)$ SG group from the $N=(4,4)$, $SO(4)\times U(1)$ 
one. We discuss various truncations and schemes of compensation of 
$N=(4,4)$, $SO(4)\times U(1)$ SG down to its $N=(4,4)$, $SU(2)$ 
superconformal descendants and, further, to different versions of 
Poincar\'e SG. A few novel possibilities are found.  
More detailed considerations with passing to component actions, etc, 
will be given elsewhere.
                                    
\setcounter{equation}{0} 
\section{Flat $SU(2)\times SU(2)$ analytic harmonic superspace} 
To proceed, we need some facts about the flat 
analytic harmonic $SU(2)\times SU(2)$ superspace. In our notation we 
will basically follow ref. \cite{ISu} with minor deviations.  

This superspace is spanned by the following set of coordinates
\be \label{anss}          
{\bf A}^{(1+2, 1+2|2,2)} = 
(z^{++}, z^{--}, \theta^{(1,0)\; 
\underline{k}\;+}, 
\theta^{(0,1)\; \underline{b}\;-}, 
u^{(\pm 1,0)}_i,\;v^{(0,\pm 1)}_a) 
\equiv (\zeta^{\mu}, u^{(\pm 1,0)}_i,\;v^{(0,\pm 1)}_a)\;. 
\ee   
Here, the  $+, -$ indices of the 
$z$ and $\theta $ coordinates are the left and right light-cone 
$SO(1,1)$ ones, while $i, \underline{k}, a, 
\underline{b}$ are doublet indices of four commuting $SU(2)$ 
groups which constitute the full automorphism 
group $SO(4)_{L}\times SO(4)_R$ of $N=(4,4)$,  $2D$ 
Poincar\'e superalgebra. In what follows we will omit the 
light-cone indices of Grassmann coordinates, keeping in mind that 
the indices $\underline{i}$ and $\underline{a}$ 
are always accompanied by the indices 
$+$ and $-$. 
The harmonic part of 
${\bf A}^{(1+2, 1+2|2,2)}$ is parametrized by two independent 
sets of harmonic variables $u^{(\pm 1,0)}_i,\;v^{(0,\pm 1)}_a$, each 
associated with one of the $SU(2)$ factors of $SO(4)_L$ and $SO(4)_R$, 
respectively (we denote these ``harmonized'' $SU(2)$ groups as 
$SU(2)_L$ and $SU(2)_R$): 
\be \label{defharm}
u^{(1,0)\;i}u^{(-1,0)}_i = 1, \;\;v^{(0,1)\;a}v_{a}^{(0,-1)} = 1\;.
\ee
The harmonics $u$ and $v$, as well as the left and right odd coordinates, 
carry two independent $U(1)$ charges ``$(n,0)$'', ``$(0,m)$'' 
which are assumed to be strictly conserved (like in the $N=2\,$, $4D$ 
harmonic superspace approach \cite{GIOS1}). 
This requirement restricts $u$ and $v$ 
to parametrize 2-spheres $SU(2)_L/U(1)_L$ and 
$SU(2)_R/U(1)_R$. The superfields given on ${\bf A}^{(1+2, 1+2|2,2)}$ 
({\it analytic} $N=(4,4)$ superfields), $\Phi^{(p,q)}(\zeta, u, v)$, are also 
labelled by a pair of such $U(1)$ charges ``$(p,q)$'' and are 
assumed to admit expansions in the double harmonic series on the 
above 2-spheres. It should be stressed that the ``harmonized'' subgroups 
$SU(2)_L, \;SU(2)_R$ and the two remaining $SU(2)$ factors of 
$SO(4)_L,\;SO(4)_R$ are realized in essentially different ways. 
Namely, the ``harmonized'' $SU(2)$ symmetries are hidden, in the sense 
that they manifest themselves only in the existence of the double 
harmonic series; on the other hand, two extra $SU(2)$ symmetries
are explicit,
as they rotate the underlined doublet indices of the analytic Grassmann 
coordinates and the related indices of component fields in the 
$\theta $ expansion of $\Phi^{(p,q)}$. Note that the latter in general 
can carry indices of any linear representation of these explicit 
$SU(2)$ symmetries.

The analytic 
superspace (\ref{anss}) is real 
with respect to the 
generalized involution ``$\sim $'' which is the product of 
ordinary complex conjugation 
and an antipodal map of the 2-spheres $SU(2)_L/U(1)_L$ 
and $SU(2)_R/U(1)_R$  
\be 
\widetilde{(\theta^{(1,0)\;\underline{i}})} = 
\theta^{(1,0)}_{\underline{i}}\;,\;\;  
\widetilde{(u^{(\pm 1,0)\;i})} = - u^{(\pm 1,0)}_i\;, 
\ee 
(and similarly for $\theta^{(0,1)\;\underline{a}},  
v^{(0,\pm 1)}_a$). 
The analytic superfields $\Phi^{(p,q)}$ 
can be chosen real with respect to this involution, provided 
$|p+q| = 2n$ 
\be 
\widetilde{(\Psi^{(p,q)})} = \Psi^{(p,q)}\;, |p+q| = 2n\;. 
\ee 

In what follows we will need the fact of the existence of 
the mutually commuting sets of harmonic derivatives   
$D^{(2,0)}$, $D_u^{(0,0)} \equiv D^0_u$ and 
$D^{(0,2)}$, $D_v^{(0,0)} \equiv D_v^0$ which preserve 
$N=(4,4)$ Grassmann harmonic analyticity, 
i.e. yield an analytic superfield when acting on some analytic superfield. 
They are given by the expressions 
\bea 
D^{(2,0)} &=& \partial^{(2,0)} + i(\theta^{(1,0)})^2 \partial_{++}\;, 
\;\;D^{(0,2)} \;=\;\partial^{(0,2)} + i(\theta^{(0,1)})^2 \partial_{--}  
\label{underfl} \\
D_u^0 &=& \partial_u^0 + \theta^{(1,0)\;\underline{i}}\frac{\partial}
{\partial \theta^{(1,0)\;\underline{i}}}\;, \;\; 
D_v^0 \;=\; \partial_v^0 + \theta^{(0,1)\;\underline{a}}\frac{\partial}
{\partial \theta^{(0,1)\;\underline{a}}}\;, 
\label{chargeop} \\
\left[\;D^0_u, D^{(2,0)}\;\right] &=& 2\;D^{(2,0)}\;,\;\; 
\left[\;D^0_v, D^{(0,2)}\;\right] \;=\; 2\;D^{(0,2)} \label{dercom}\;.
\eea
Here $\partial_{\pm\pm} = \partial /\partial z^{\pm\pm}$ and 
\be
\partial^{(2,0)} = u^{(1,0)\;i} \frac{\partial}{\partial u^{(-1,0)\;i}}\;, \;\;
\partial^{0}_u = u^{(1,0)\;i} \frac{\partial}{\partial u^{(1,0)\;i}} - 
u^{(-1,0)\;i} \frac{\partial}{\partial u^{(-1,0)\;i}}\;, 
\ee
(the same formulas are valid for $\partial^{(0,2)}$ and $\partial^{0}_v$ 
with the change 
$u\rightarrow v$). The operators $D_u^0$, $D_v^0$ count the $U(1)$
charges of the analytic superfields 
\be
D^0_u\, \Phi^{(p,q)}(\zeta, u, v) = p\,\Phi^{(p,q)}(\zeta, u, v)\;,\;\;
D^0_v\, \Phi^{(p,q)}(\zeta, u, v) = q\, \Phi^{(p,q)}(\zeta, u, v)\;.
\ee

In the analytic superspace (\ref{anss}) one can realize 
two different infinite-dimensional groups of superconformal 
transformations. Each group consists of two commuting light-cone branches, 
the left and right ones, having as the algebra the classical 
$N=4$, $SU(2)$ superconformal algebra (SCA) \cite{Ademollo}. Without loss of  
generality we can specialize, e.g., to the left sector. 
It turns out that the form of the relevant superconformal 
transformations is basically specified by 
the transformation law of the analyticity-preserving 
covariant harmonic derivative $D^{(2,0)}$ (or $D^{(0,2)}$ in the right 
sector). 

The basic distinguishing feature of the first group is that 
it does not touch the harmonics 
\be  \label{tranharII}
\delta_{I}\;u^{(\pm 1,0)}_i = 0\;.
\ee
Its realization in ${\bf A}^{(1+2,1+2|2,2)}$ is completely fixed by 
the requirement that $D^{(2,0)}$ is invariant
\be \label{scg2}
\delta_{I}\;D^{(2,0)} = 0\;.
\ee

The second superconformal group has the same Lie bracket 
structure as the first one, but it acts on {\it all} the left
coordinates of 
${\bf A}^{(1+2,1+2|2,2)}$,
including the harmonic ones $u^{(\pm 1,0)}$. We 
give here only the generic form of transformations of harmonics and 
the derivative $D^{(2,0)}$  \cite{ISu}
\bea 
\delta_{II}\; u^{(1,0)}_i &=& \Lambda^{(2,0)}_I (z^{++}, \theta^{(1,0)}, u) 
u^{(-1,0)}_i\;,\; 
\delta_{II}\; u^{(-1,0)}_i \;=\; 0 \nonumber \\
\delta_{II}\; D^{(2,0)} &=& - \Lambda^{(2,0)} D^0_u\;,\;\; 
\Lambda^{(2,0)} \;=\; D^{(2,0)} \Lambda_L\;,\;\;  
D^{(2,0)} \Lambda^{(2,0)} \;=\; 0 \;. \label{scg1}
\eea

The main difference between these two $N=4$, $SU(2)$ superconformal 
groups lies in the realization of their affine $SU(2)$ subgroups: 
the second one acts on the indices $i,j$ and affects both the 
Grassmann and harmonic coordinates, while the first one acts only 
on the underlined indices and so affects only $\theta $'s. These groups 
do not commute; their closure is the ``large'' $N=4$, $SO(4)\times U(1)$ 
group \cite{{Ademollo},{Belg2},{IKLev1},{IKLev2}}. For our further purposes 
it will be important that the latter involves an extra $U(1)$ 
affine (Kac-Moody) symmetry 
with the dimensionless holomorphic parameter $\lambda_L(z^{++})$ 
(or $\lambda_R(z^{--})$ in the right sector). 
It is realized, e.g., on $u^{(1,0)\;i}$ as
\cite{ISu}  
\be  \label{u1km}
\delta_{U(1)} u^{(1,0)\;i} = 
(\,D^{(2,0)} \lambda_L(z)\,)\, u^{(-1,0)\;i} = i(\theta^{(0,1)})^2 
\partial_{++}\lambda_L(z)\, u^{(-1,0)\;i}\;.
\ee
The ``large'' superconformal algebra corresponds to the most general 
solution \cite{delsok} of the 
constraints on $\Lambda^{(2,0)}$ in eq. \p{scg1}, while two of
its $SU(2)$ subalgebras (SCA-I and SCA-II in what follows) 
are singled out by some additional conditions. 
Here we will not present the 
explicit form of the coordinate transformations of all these 
superconformal groups (see \cite{{ISu},{IS3}} for details),
since we will recover them as flat limits of the appropriate SG groups 
in the next Sections. Notice the following important 
property: both $N=4$, $SU(2)$ superconformal groups, and hence 
their closure, leave invariant the analytic superspace 
integration measure 
$\mu^{(-2,-2)} = d^2z d^2\theta^{(1,0)} d^2\theta^{(0,1)} 
[du] [dv]$:
\be  \label{measI,II}
\delta_I\; \mu^{(-2,-2)} = \delta_{II}\; \mu^{(-2,-2)} = 0\;.
\ee    

The last topic of this introductory Section is the 
harmonic superspace description of some important $N=(4,4)$ multiplets.
We start with one of the possible $N=(4,4)$ twisted 
chiral multiplets \cite{{GHR},{IK1}}, namely, the one having a simple 
description in $SU(2)\times SU(2)$ harmonic analytic superspace. 
It is represented by a real analytic $(4,4)$ superfield 
$q^{(1,1)}(\zeta, u,v)$ subject to 
the constraints 
\be
D^{(2,0)}q^{(1,1)} = D^{(0,2)}q^{(1,1)} = 0\;.
\label{qucons}
\ee
They leave in $q^{(1,1)}$ $8+8$ independent components \cite{ISu}, 
just the off-shell field content of $N=(4,4)$ twisted 
multiplet. The superfield $q^{(1,1)}$ is scalar with respect to 
the first $N=4$, $SU(2)$ superconformal group but it is 
transformed with the weight 1 under the second one (this is necessary for 
preserving the constraints (\ref{qucons}))
\be 
\delta_{I}\; q^{(1,1)} = 0 \label{tranqu} \;, \quad
\delta_{II}\; q^{(1,1)} = \Lambda_L\, q^{(1,1)}
\ee
(the transformations from the right-handed branches are similar). The 
physical dimension components of $q^{(1,1)}$ 
(four dimension 0 bosons and eight dimension 1/2 fermions) behave in 
different ways under these two kinds of 
$N=(4,4)$, $SU(2)$ transformations. In particular, the $SU(2)$ affine 
transformations from the first superconformal group act only on 
fermions, while those from the second group act 
both on bosons and fermions. The physical bosonic fields are naturally 
combined, with respect to the latter transformations and 
their right-handed counterparts, into a 
2$\times$2 matrix $q^{ia}(z^{++},z^{--})$ on which the left 
(right) conformal $SU(2)$ acts as a
left (right) multiplication . So the purely $SU(2)$ part of 
$q^{ia}$ represents the coset $SU(2)_L\times SU(2)_R/SU(2)_{diag}$, and it 
is not too surprising that the $q^{(1,1)}$ action invariant under 
the second superconformal group is none other than $N=(4,4)$ extension
of the  $SU(2)$ WZW sigma model action. Indeed, it is
just the $N=4$, $SU(2)\times U(1)$ 
WZW sigma model action of ref.
\cite{{IK1},{IKLev1},{IKLev2},{belg},{RSS},{GorI}}. 
The $SU(2)\times SU(2)$ analytic superspace form of this action
reads \cite{ISu} 
\be \label{wzwact} 
S_{wzw} = -\frac{1}{4\gamma^2} \int \mu^{(-2,-2)} \;\hat{q}^{(1,1)} 
\hat{q}^{(1,1)} 
\left( \frac{1}{(1+X)X} -\frac{\mbox{ln}(1+X)}{X^2} \right)\;, 
\label{confact} 
\ee 
where  
\be \label{defc11}
\hat{q}^{(1,1)} = q^{(1,1)} - c^{(1,1)}\;,\;X = c^{(-1,-1)}
\hat{q}^{(1,1)}\;,\; 
c^{(\pm 1,\pm 1)} = c^{ia}u^{(\pm1,0)}_iv^{(0,\pm1)}_a \;,\; 
c^{ia}c_{ia} = 2\;, 
\ee
and $\gamma$ is a dimensionless sigma model coupling constant. 
Despite the presence of an extra quartet constant $c^{ia}$ in the 
analytic superfield Lagrangian, the action actually does not 
depend on $c^{ia}$ \cite{ISu}. 

We wish to stress that the action (\ref{wzwact}) is unique (up to 
adding full harmonic derivatives) in the sense 
that it is the only possible action of a single superfield $q^{(1,1)}$ 
invariant under the second $N=(4,4),\, SU(2)$ superconformal group. As we 
will see later, in the 
curved case the superfield $q^{(1,1)}$ serves as a compensator 
which breaks the appropriate $N=(4,4)$, $SU(2)$ SG group 
(having as the rigid limit the second $N=(4,4)$, $SU(2)$ superconformal 
group) down to the supergroup of minimal $N=(4,4)$, $2D$ SG \cite{Gates1}. 

As for the first superconformal group, an arbitrary 
action of the superfield $q^{(1,1)}$, 
\be 
S_q = \int \mu^{(-2,-2)}\; {\cal L}^{(2,2)} (q^{(1,1)\;M} 
(\zeta,u,v),\; 
u,\; v)~, 
\label{genaction} 
\ee 
is invariant with respect to it. As a consequence of this property, 
the particular $q^{(1,1)}$ action (\ref{wzwact}) is invariant under 
both
$N=(4,4)$, $SU(2)$ superconformal groups and, hence, under their closure, i.e.
the ``large'' $N=(4,4)$, $SO(4)\times U(1)$ superconformal group. Note that 
$q^{(1,1)}$ transforms under the left affine $U(1)$ 
transformations (\ref{u1km}) as 
\be \label{KMqu}
\delta_{U(1)} q^{(1,1)} = \lambda_L(z^{++})\, q^{(1,1)}\;
\ee
(and analogously under their right-handed counterparts). The full 
transformation law of $q^{(1,1)}$ under the left ``large'' group 
looks like the second law in eq. 
\p{tranqu}, with $\Lambda_L = \lambda_L(z^{++}) + 
\lambda^{(ik)}(z^{++})u^{(1,0)}_iu^{(-1,0)}_k + ... $. Further details 
will be given in Sect. 4. It is worth mentioning that the general action 
\p{genaction} always yelds the sigma model {\it with torsion} in the sector of
physical bosons, just of the same kind as in the $N=(4,4)$
supersymmetric subclass of general $N=(2,2)$ chiral and twisted chiral
superfield sigma models explored in \cite{GHR}. The actions of other matter
multiplets in $SU(2)\times SU(2)$ harmonic superspace reveal the same
characteristic feature. This is the radical difference of the considered case 
from the dimensionally-reduced off-shell sigma model actions of
hypermultiplets in the standard harmonic superspace with one set of the
$SU(2)$ harmonic variables \cite{GIOS1}: for physical bosons they yield the
{\it torsionless} hyper-K\"ahler sigma model actions.   

Note that there exist other types of twisted $N=4$ multiplets, 
with the same number of off-shell components, but with different  
realizations of various $SU(2)$ factors of the full $SO(4)_L\times 
SO(4)_R$ automorphism group of rigid $N=(4,4), \;\;2D$ SUSY 
\cite{{gake},{IS2}}. 
Respectively, the above two $N=(4,4)$, $SU(2)$ superconformal 
groups are realized in different ways on these multiplets. In particular, 
there exists a sort of twisted multiplet 
on which the first and second 
superconformal groups act in the way just opposite to their action on 
$q^{(1,1)}$.
\footnote{In \cite{{Gates3},{gake}} such a multiplet 
is called TM-I as opposed to $q^{(1,1)}$ which 
is TM-II in this classification. Such a classification makes sense 
with respect to a fixed $N=(4,4),\, SU(2)$ SCA: if the affine 
$SU(2)_L\times SU(2)_R$ subgroup acts 
both on the physical bosons and fermions, one deals with TM-II,
whereas if it acts 
only on fermions, one faces TM-I. Conversely, $q^{(1,1)}$ is TM-I 
with respect to the
first of the two $N=(4,4),\, SU(2)$ SCAs defined above, but 
it is TM-II with respect to the second one.} 

The $SU(2)\times SU(2)$ harmonic superspace description of 
these complementary twisted multiplets \cite{IS2} is somewhat 
more complicated. Nevertheless, all of them 
can be coupled to the $N=(4,4)$ Beltrami-Weyl SG multiplets to be defined 
below and so can serve as compensators. We are planning to present  
these couplings in a future work.   

Finally, we mention one more analytic $SU(2)\times SU(2)$ harmonic 
supermultiplet which will be used in Sect. \ref{supmat} as a compensator 
reducing the
$N=(4,4)$, $SO(4)\times U(1)$ SG group to one of its $N=(4,4)\;$, $SU(2)$ 
subgroups. It is represented by a pair 
of analytic superfields $N^{(2,0)}$, $N^{(0,2)}$ satisfying 
the constraints \cite{IS3}
\bea
D^{(2,0)}N^{(2,0)} + N^{(2,0)}N^{(2,0)} &=& 0,\;\; 
D^{(0,2)}N^{(0,2)} + N^{(0,2)}N^{(0,2)} = 0,\nn \\
D^{(2,0)}N^{(0,2)} - D^{(0,2)}N^{(2,0)} &=& 0~.  \label{flatQconstr}
\eea
These constraints are analogous to those defining the so-called nonlinear 
supermultiplet \cite{holl} in the $N=2$, $4D$ harmonic superspace 
(the latter goes into $N=(4,4)$, 
$SU(2)_{diag}$ harmonic superspace upon reduction to $2D$). 
They are obviously covariant under the first $N=(4,4)\;$, $SU(2)$ 
superconformal group, if $N^{(2,0)}, N^{(0,2)}$ are assumed to 
transform as scalars with respect 
to it. They are also covariant under the second group, provided 
$N^{(2,0)}, N^{(0,2)}$ transform according to 
\be \label{transfQ}
\delta_{II} N^{(2,0)} =
\Lambda^{(2,0)}, \;\;\;\delta_{II} N^{(0,2)} = \Lambda^{(0,2)}\;.
\ee
The simplest invariant action (with the correct sign of the
kinetic terms of the physical fields) is as follows: 
\be \label{lagrN}
S_N \sim - \int \mu^{(-2,-2)}\, N^{(2,0)}N^{(0,2)}\;.
\ee
To see that it is invariant (up to surface terms) under 
\p{transfQ}, one should take into account the invariance of the analytic 
superspace integration measure and the properties 
\be
\Lambda^{(2,0)} = D^{(2,0)}\Lambda_L,\;\;
\Lambda^{(0,2)} = D^{(0,2)}\Lambda_R, \;\; D^{(2,0)}\Lambda_R = 
D^{(0,2)}\Lambda_L = 0\;.
\ee
The pair $N^{(2,0)}, N^{(0,2)}$ describes $32 + 32$ off-shell 
degrees of freedom and is dual-equivalent to four $q^{(1,1)}$ 
superfields \cite{IS3}.    

Having the multiplet $N^{(2,0)}, N^{(0,2)}$, one can define further 
consistent non-linear multiplets $G^{(2,0)}, G^{(0,2)}$ which 
are zero-weight scalars under both $N=(4,4),\, SU(2)$ superconformal groups 
\be  \label{rigG1}
\delta_{I,II}\, G^{(2,0)} = \delta_{I,II}\, G^{(0,2)} = 0~.
\ee  
The corresponding constraints (covariant with respect to both superconformal 
groups) are a slight modification of \p{flatQconstr}
\bea
(D^{(2,0)} + 2N^{(2,0)})G^{(2,0)} + \alpha \,G^{(2,0)}G^{(2,0)} &=& 0~, \nn \\
(D^{(0,2)}+ 2N^{(0,2)})G^{(0,2)} + \alpha \,G^{(0,2)}G^{(0,2)} &=& 0~,\nn \\
D^{(2,0)}G^{(0,2)} - D^{(0,2)}G^{(2,0)} &=& 0~, \label{flGconstr}
\eea 
where $\alpha $ is an arbitrary dimensionless parameter (it can be equal 
to zero). All such representations comprise $32 + 32$ off-shell
degrees of freedom. Their Lagrangians are bilinears like in \p{lagrN}. 

\setcounter{equation}{0}     
\section{Curved $SU(2)\times SU(2)$ analytic superspace and \break 
N=(4,4) Beltrami-Weyl multiplet} 
By analogy with the $N=2$, $4D$ case \cite{{GIKOS},{GIOS2}} we assume 
that the fundamental group of $N=(4,4)\;$, $2D$ conformal supergravity is 
represented by the following diffeomorphisms of the analytic 
harmonic $SU(2)\times SU(2)$ superspace 
\bea 
&&\delta \zeta^\mu = \Lambda^\mu (\zeta, u,v),\;
\delta u^{(1,0)}_i = \Lambda^{(2,0)}(\zeta, u,v) u^{(-1,0)}_i,\; 
\delta v^{(0,1)}_a = \Lambda^{(0,2)}(\zeta, u,v) v^{(0,-1)}_a,  \nonumber \\
&& \delta u^{(-1,0)}_i = \delta v^{(0,-1)}_a =0\;. \label{sgg}
\eea  
Here $\zeta^\mu = (z^{++}, z^{--}, \theta^{(1,0)\; 
\underline{k}\;+}, \theta^{(0,1)\; \underline{b}\;-})$ as in 
\p{anss} and the gauge parameters $\Lambda^\mu$, $\Lambda^{(2,0)}$, 
$\Lambda^{(0,2)}$ are arbitrary functions 
over the {\it whole} harmonic analytic superspace 
${\bf A}^{(1+2,1+2|2,2)}$. These transformation laws preserve 
the defining relations of harmonic 
variables (\ref{defharm}) and the reality of ${\bf A}^{(1+2,1+2|2,2)}$ 
with respect to the ``$\sim{ }$'' conjugation. The 
analyticity-preserving harmonic derivatives $D^{(2,0)}$ and $D^{(0,2)}$ are 
covariantized by introducing appropriate analytic vielbeins 
\bea 
D^{(2,0)} \Rightarrow  \nabla^{(2,0)} &=& D^{(2,0)} + H^{(2,0)\;\mu} 
\partial_\mu 
+ H^{(4,0)}\partial^{(-2,0)} + 
H^{(2,2)}\partial^{(0,-2)} \nonumber \\
&\equiv & D^{(2,0)} +H^{(2,0)\;M}\partial_M \;, 
\nonumber \\
D^{(0,2)} \Rightarrow  \nabla^{(0,2)} &=& 
D^{(0,2)} + H^{(0,2)\;\mu}\partial_\mu 
+ \tilde{H}^{(2,2)}\partial^{(-2,0)} + H^{(0,4)}\partial^{(0,-2)} \nonumber \\
&\equiv & D^{(0,2)} +H^{(0,2)\;M}\partial_M \;,   
\label{covdII}
\eea
where we used the notation 
\bea
&& M = (\mu, (2,0), (0,2)), \quad \partial_M = (\partial_\mu, 
\partial^{(-2,0)},\partial^{(0,-2)}),\nonumber \\ 
&& \partial^{(-2,0)} = u^{(-1,0)\;i}
\frac{\partial}{\partial u^{(1,0)\;i}}, \quad 
\partial^{(0,-2)} = v^{(0,-1)\;a}
\frac{\partial}{\partial v^{(0,1)\;a}} 
\eea
and separated the flat parts of the vielbein components in 
front of $\partial_{++}$ in $\nabla^{(2,0)}$ and $\partial_{--}$ 
in $\nabla^{(0,2)}$. In eqs. (\ref{covdII}) all the vielbeins are 
analytic $N=(4,4)$, $2D$ 
superfields, 
$$
H^{(2,0)\;M}=H^{(2,0)\;M}(\zeta,u,v)\;,\;\;
H^{(0,2)\;M}=H^{(0,2)\;M}(\zeta,u,v)\;.
$$ 
The flat limit is achieved by putting 
them equal to zero.
The $U(1)$ charge-counting operators $D^0_u$ and $D^0_v$ retain their 
flat form (\ref{chargeop}).

Again in analogy with refs. \cite{{GIKOS},{GIOS2}}, we postulate for 
$\nabla^{(2,0)}$, $\nabla^{(0,2)}$ the following transformation law under the 
$N=(4,4)$ SG group (\ref{sgg})
\be \label{tranD}
\delta \nabla^{(2,0)} = -\Lambda^{(2,0)} D^0_u\;,\;\;\; 
\delta \nabla^{(0,2)} = -\Lambda^{(0,2)} D^0_v\;,
\ee
whence 
\bea 
\delta H^{(2,0)\;++} &=& \nabla^{(2,0)} \Lambda^{++} - 2i 
\Lambda^{(1,0)}\theta^{(1,0)}\;,\;\; 
\delta H^{(2,0)\;--} \;=\; \nabla^{(2,0)} \Lambda^{--}\:,  
\nonumber \\ 
\delta H^{(3,0)\;\underline{i}} &=& \nabla^{(2,0)} 
\Lambda^{(1,0)\;\underline{i}} 
- \Lambda^{(2,0)}\theta^{(1,0)\;\underline{i}}\;, \;\;
\delta H^{(2,1)\;\underline{a}} \;=\; 
\nabla^{(2,0)} \Lambda^{(0,1)\;\underline{a}}\;, \nonumber \\
\delta H^{(4,0)} &=& \nabla^{(2,0)}\Lambda^{(2,0)}\;,\;\; 
\delta H^{(2,2)} \;=\; \nabla^{(2,0)}\Lambda^{(0,2)}\;,
\label{hdir20} \\
\delta H^{(0,2)\;++} &=& \nabla^{(0,2)} \Lambda^{++}  
\;, \;\;
\delta H^{(0,2)\;--} \;=\; \nabla^{(0,2)} \Lambda^{--}  
-2i \Lambda^{(0,1)}\theta^{(0,1)}\;, \nonumber \\
\delta H^{(1,2)\;\underline{i}} &=& 
\nabla^{(0,2)} \Lambda^{(1,0)\;\underline{i}}\;, 
\;\;
\delta H^{(0,3)\;\underline{a}} \;=\; \nabla^{(0,2)} 
\Lambda^{(0,1)\;\underline{a}} - \Lambda^{(0,2)} 
\theta^{(0,1)\;\underline{a}}\;, 
\nonumber \\
\delta \tilde{H}^{(2,2)} &=& \nabla^{(0,2)}\Lambda^{(2,0)}\;,\;\; 
\delta H^{(0,4)} \;=\; \nabla^{(0,2)}\Lambda^{(0,2)}\;. 
\label{hdir02}
\eea 
 From now on, the similarity with the $N=2$, $4D$ construction 
ceases to be literal and the
specificity of the $N=(4,4)$ case comes into play.

First of all, we wish to generalize the notion of the twisted 
analytic superfield $q^{(1,1)}$ to 
the curved case and hence need to find a correct generalization of the 
defining constraints (\ref{qucons}) and the superconformal transformation 
laws (\ref{tranqu}). As we have started with the most general diffeomorphism 
group of the analytic superspace, we expect it to yield, in the flat limit, 
the product of the left and right ``large''  $SO(4)\times U(1)$ 
superconformal groups, including their $U(1)$ affine subgroups 
with the parameters $\lambda_L(z^{++})$, $\lambda_R(z^{--})$. 
However, a close inspection of the analytic superfield gauge parameters 
$\Lambda^\mu(\zeta, u,v)$, $\Lambda^{(2,0)}(\zeta,u,v)$ and 
$\Lambda^{(0,2)}(\zeta,u,v)$ shows that there is no place in them for 
such dimensionless parameters (these can appear only with their $z$ 
derivatives). To generalize 
the transformation laws of $q^{(1,1)}$ (\ref{tranqu}), (\ref{KMqu}) to the 
curved case, we are 
led to introduce two extra {\it independent} analytic gauge functions 
$$
\Lambda_L(\zeta, u,v) = \lambda_L(z^{++},z^{--}) + ... \;,\;\; 
\Lambda_R(\zeta, u,v) = \lambda_R (z^{++},z^{--}) + ...
$$
and to ascribe the following transformation laws to $q^{(1,1)}$ 
\be \label{qutranloc}
\delta q^{(1,1)} = (\Lambda_L + \Lambda_R) q^{(1,1)}\;.
\ee
We call these transformations the ``$U(1)$ weight'' ones, to distinguish 
them from the harmonic $U(1)$ phase transformations. We normalize 
the left and right $U(1)$ weights $J_L$ and $J_R$ as
\be \label{quweight}
J_L q^{(1,1)} = J_R q^{(1,1)} = q^{(1,1)}\;.
\ee
At this stage, the $U(1)$ weight analytic parameters
$\Lambda_L$, $\Lambda_R$ 
are entirely unrelated to those of the coordinate transformations. 

Such a relation naturally comes out, as a result of choosing 
the appropriate transformation law for the $U(1)$ weight-covariantized 
harmonic derivatives and fixing a proper gauge. 

We covariantize $\nabla^{(2,0)}$, $\nabla^{(0,2)}$ by introducing 
four analytic superfield $U(1)$ connections 
$H^{(2,0)}_L(\zeta,u,v)$, 
$H^{(2,0)}_R(\zeta,u,v)$, $H^{(0,2)}_L(\zeta,u,v)$, 
$H^{(0,2)}_R(\zeta,u,v)$
\bea
\nabla^{(2,0)} &\Rightarrow & {\cal D}^{(2,0)} = 
\nabla^{(2,0)} +  H^{(2,0)}_L J_L + H^{(2,0)}_R J_R \nonumber \\ 
\nabla^{(0,2)} &\Rightarrow & {\cal D}^{(0,2)} = 
\nabla^{(0,2)} +  H^{(0,2)}_L J_L + H^{(0,2)}_R J_R \;,
\label{u1covd}
\eea
and postulate the following transformation laws for ${\cal D}^{(2,0)}$, 
${\cal D}^{(0,2)}$
\bea 
\delta {\cal D}^{(2,0)} &=& -\Lambda^{(2,0)} (D^0_u - J_L) - 
\nabla^{(2,0)}\Lambda_L\;J_L - \nabla^{(2,0)}\Lambda_R\;J_R\;, \nonumber \\ 
\delta {\cal D}^{(0,2)} &=& -\Lambda^{(0,2)}\; (D^0_v - J_R)-
\nabla^{(0,2)}\Lambda_L\;J_L - \nabla^{(0,2)}\Lambda_R\;J_R \;. 
\label{tranu1D}
\eea
The transformation laws of the vielbeins in $\nabla^{(2,0)}$, 
$\nabla^{(0,2)}$ do not change, 
while the newly introduced $U(1)$ connections are transformed as 
\bea  
\delta H^{(2,0)}_L &=& \Lambda^{(2,0)} - \nabla^{(2,0)}\Lambda_L \;, \;\;
\delta H^{(2,0)}_R \;=\; - \nabla^{(2,0)}\Lambda_R \;, \nonumber \\
\delta H^{(0,2)}_L &=& - \nabla^{(0,2)}\Lambda_L \;,
\;\;\delta H^{(0,2)}_R \;=\; \Lambda^{(0,2)} - \nabla^{(0,2)}\Lambda_R \;.
\label{tranu1con}
\eea
The ${\cal D}^{(2,0)}$ 
and ${\cal D}^{(0,2)}$ derivatives of the analytic 
superfield  $\Phi^{(p,q)}$, with the left and right $U(1)$ weights 
equal to $l$ and $r$, are transformed as follows: 
\bea 
\delta {\cal D}^{(2,0)} \Phi^{(p,q)} &=& -\Lambda^{(2,0)}(p-l)\Phi^{(p,q)} + 
(l \Lambda_L + r \Lambda_R) {\cal D}^{(2,0)} \Phi^{(p,q)}\;, \nonumber \\
\delta {\cal D}^{(0,2)} \Phi^{(p,q)} &=& -\Lambda^{(0,2)}(q-r)
\Phi^{(p,q)} + 
(l \Lambda_L + r \Lambda_R) {\cal D}^{(0,2)} \Phi^{(p,q)} \;. \label{tranpqph}
\eea
We see that only provided $p=l$, $q=r$, these derivatives are actually 
covariant, i.e. they transform as the superfield $\Phi^{(p,q)}$ itself. 
But this is precisely what happens for $q^{(1,1)}$, which possesses 
$J_L = J_R =1$. Therefore, 
as the appropriate curved generalization of the constraints (\ref{qucons}), 
we choose the following ones:
\bea
{\cal D}^{(2,0)} q^{(1,1)} &=& (\nabla^{(2,0)} + 
H^{(2,0)}_L + H^{(2,0)}_R) q^{(1,1)} 
\;=\; 0 \;, \nonumber \\
{\cal D}^{(0,2)} q^{(1,1)} &=& (\nabla^{(0,2)} + 
H^{(0,2)}_L + H^{(0,2)}_R) q^{(1,1)} 
\;=\; 0 \;. \label{curqucons} 
\eea

Before going further, let us adduce some reasoning in favor of 
the choice of the transformation laws of 
${\cal D}^{(2,0)}$, ${\cal D}^{(0,2)}$ in the form (\ref{tranu1D}). 
The primary reason for this choice is the desire to relate  
the coordinate transformations with the $U(1)$ weight 
transformations, so as to eventually ensure a correct flat limit. 
Indeed, from eqs. (\ref{tranu1con}) it follows 
that the connections $H^{(2,0)}_L$, $H^{(0,2)}_R$ can be entirely 
gauged away, thereby establishing the sought relation  
\be  \label{basgauge}
H^{(2,0)}_L = H^{(0,2)}_R = 0 \Rightarrow 
\Lambda^{(2,0)} = \nabla^{(2,0)}\Lambda_L\;,
\;\; \Lambda^{(0,2)} = \nabla^{(0,2)}\Lambda_R \;.
\ee
In what follows we will frequently stick to this gauge. 
One more argument why we should assume (\ref{tranu1D}) is based 
on an analogy with the harmonic space 
description of quaternionic manifolds in \cite{GIO}. There, the 
analyticity-preserving harmonic derivative in the analytic basis 
necessarily 
involves an analytic connection $\phi^{++}$ associated with the so 
called ``$Sp(1)$ weight''. Its transformation law literally 
mimics that of $H^{(2,0)}_L$, $H^{(0,2)}_R$, so it is natural to 
assume that the $U(1)$ weights $J_L$, $J_R$ and the associated 
analytic superfield parameters $\Lambda_L$ and $\Lambda_R$ are direct 
analogs of 
the just mentioned $Sp(1)$ weight and the related analytic parameter inherent
to the quaternionic manifolds \footnote{A deep analogy between 
the description of quaternionic manifolds in the harmonic space and that of 
conformal $N=2, \;4D$ SG in the harmonic superspace was pointed 
out in \cite{GIO}.}. Of course, the most direct way 
to justify the transformation law (\ref{tranu1D}) would be to deduce it 
proceeding from the appropriate constraints in the standard 
$N=(4,4)$ superspace. An alternative way is to show that it leads to a 
self-consistent SG theory, still in the framework of the analytic 
superspace. This is just what we are going to demonstrate.

An important consequence of the presence of two independent harmonic 
constraints in the definition of the twisted superfield $q^{(1,1)}$, eqs. 
(\ref{curqucons}), is the integrability condition 
\be  \label{intqu}
[{\cal D}^{(2,0)}, {\cal D}^{(0,2)}]q^{(1,1)} =0 \;.
\ee 
It is easy to see that the direct generalization of the flat 
condition $[D^{(2,0)}, D^{(0,2)}]=0$, namely, 
$$
[{\cal D}^{(2,0)}, {\cal D}^{(0,2)}]=0\;,
$$ 
is not covariant under (\ref{tranu1D}). The covariant version of this 
constraint is as follows:
\be \label{constrD}
[{\cal D}^{(2,0)}, {\cal D}^{(0,2)}] = 
- H^{(2,2)}(D^0_v - J_R) + \tilde{H}^{(2,2)} 
(D_u^0 - J_L)~.
\ee
It is evident that eq. (\ref{intqu}) is automatically satisfied as 
a consequence of (\ref{constrD}) and \p{quweight}. This constraint implies
\bea
\nabla^{(2,0)}H^{(0,2)}_L - \nabla^{(0,2)}H^{(2,0)}_L + \tilde{H}^{(2,2)} 
&=& 0~, \nonumber \\
\nabla^{(2,0)}H^{(0,2)}_R - \nabla^{(0,2)}H^{(2,0)}_R - H^{(2,2)} &=& 0 
\label{constrh22}
\eea
and 
\be \label{consnab}
[\nabla^{(2,0)}, \nabla^{(0,2)}] = - H^{(2,2)} D^0_v + 
\tilde{H}^{(2,2)} D^0_u\;.
\ee
 From the latter relation one deduces the constraints on the analytic 
vielbeins
\bea
\nabla^{(2,0)} H^{(0,2)\;++} - \nabla^{(0,2)} H^{(2,0)\;++} - 2i 
H^{(1,2)} \theta^{(1,0)} &=& 0 \;,\nonumber \\
\nabla^{(2,0)} H^{(0,2)\;--} - 
\nabla^{(0,2)} H^{(2,0)\;--} + 2i H^{(2,1)}\theta^{(0,1)} 
&=& 0 \;, \nonumber \\
\nabla^{(2,0)} H^{(1,2)\;\underline{i}} - \nabla^{(0,2)} 
H^{(3,0)\;\underline{i}} - \tilde{H}^{(2,2)} 
\theta^{(1,0)\;\underline{i}}&=& 0 
\;, \nonumber \\
\nabla^{(2,0)} H^{(0,3)\;\underline{a}} - \nabla^{(0,2)} 
H^{(2,1)\;\underline{a}} 
+ H^{(2,2)}\theta^{(0,1)\;\underline{a}} &=& 0 \;, 
\nonumber \\
\nabla^{(2,0)} H^{(0,4)} - \nabla^{(0,2)} H^{(2,2)} &=& 0\;, \nonumber \\
\nabla^{(2,0)} \tilde{H}^{(2,2)} - \nabla^{(0,2)} H^{(4,0)} &=& 0\;. 
\label{constrH}
\eea
Thus we see that in the $N=(4,4), \;SU(2)\times SU(2)$ case the
analytic vielbeins and $U(1)$ 
connections covariantizing $D^{(2,0)}$, $D^{(0,2)}$ are necessarily 
constrained. This is the crucial difference from the formulation of 
$N=2$, $4D$ conformal SG in the standard 
harmonic superspace \cite{{GIKOS},{GIOS2}}, where the analogous quantities 
are unconstrained analytic superfields, i.e. the prepotentials of the
theory. 
Of course, this peculiarity is a direct consequence of the presence of two 
independent sets of harmonic variables in the considered case. 

For the time being, we do not know how to solve 
(\ref{constrh22}), (\ref{constrH}) via unconstrained superfield 
prepotentials. To single out the irreducible 
field representation carried by vielbeins and $U(1)$ connections, 
we keep to another strategy. 
Namely, we use the initial gauge freedom to gauge away from these objects 
as many components as possible, then substitute the resulting 
expressions into the constraints 
and solve the latter in this WZ-type gauge. Eventually, it turns out that 
the solution exists, is unique and is not reduced to a pure gauge. 
The superfield constraints prove to be purely kinematic: indeed,
they do not imply any differential conditions, nor 
equations of motion, for the remaining fields. At present we are aware of 
the full nonlinear solution 
of these constraints. Here, we limit 
ourselves to the linearized level. This is quite sufficient  
for revealing the irreducible field contents of the SG theory under 
consideration. 

In the present case, one can choose the WZ gauge in several different ways, 
the basic criterion for one or another choice being the desire to 
simplify the constraints (\ref{constrh22}), (\ref{constrH}) as much as 
possible. As a first step, 
we choose the gauge (\ref{basgauge}) and the following additional ones
\bea 
H^{(2,0)\;++} &=& H^{(0,2)\;--} \;=\; H^{(3,0)\;\underline{i}} \;=\; 
H^{(0,3)\;\underline{a}} \;=\; 0 \;,\label{gauge1} \\
H^{(4,0)} &=& H^{(0,4)} \;=\; 0\;. \label{gauge2}
\eea
These gauges restrict in a certain way the original gauge parameters.
At the 
considered linearized level, (\ref{gauge1}) and (\ref{gauge2}) give rise to 
the following relations:
\bea
D^{(2,0)}\Lambda^{++} - 2i\Lambda^{(1,0)}\theta^{(1,0)} &=& 0\;,\; 
D^{(2,0)}\Lambda^{(1,0)\;\underline{i}} - D^{(2,0)}\Lambda_L \;
\theta^{(1,0)\;\underline{i}} \;=\; 0\;, \nonumber \\
D^{(0,2)}\Lambda^{--} - 2i\Lambda^{(0,1)}\theta^{(0,1)} &=& 0\;, \;
D^{(0,2)}\Lambda^{(0,1)\;\underline{a}} - D^{(0,2)}\Lambda_R \;
\theta^{(0,1)\;\underline{a}} \;=\; 0\;, \nonumber \\
(D^{(2,0)})^2 \Lambda_L \;=\; (D^{(0,2)})^2 \Lambda_R &=& 0\;,\label{rel1} 
\eea
which strictly fix the $u$ or $v$ dependence of the relevant parameters 
(depending on which derivative, i.e. either $D^{(2,0)}$ or
$D^{(0,2)}$, enters 
the given relation). After this, there still remains a freedom 
associated with the surviving harmonic dependence. This freedom can be
used to further gauge away some of 
the components in the double harmonic expansion of the remaining 
vielbeins $H^{(2,0)\;--}$, $H^{(0,2)\;++}$, $H^{(2,1)\;\underline{a}}$, 
$H^{(1,2)\;\underline{i}}$ and the $U(1)$ connections $H^{(2,0)}_R$, 
$H^{(0,2)}_L$. At this stage, the $u$ and $v$ dependence of all analytic 
superfield gauge parameters is completely fixed and we are left with a finite 
set of the component parameters. However, in the vielbeins and connections 
one still finds a non-trivial harmonic dependence which is entirely 
fixed only after imposing the constraints. The final 
expressions for the vielbeins, connections and superfield gauge 
parameters at the linearized level are as follows: 
\bea
H^{(2,0)\;--} &=& i(\theta^{(1,0)})^2 \{ h^{--}_{++} - 2i 
\theta^{(0,1)}_{\;\underline{a}} h^{-\;a\underline{a}}_{++} 
v^{(0,-1)}_a -i 
(\theta^{(0,1)})^2 h^{(ab)}_{++} v^{(0,-1)}_av^{(0,-1)}_b \}\;, \nonumber \\
H^{(2,1)\;\underline{a}} &=& i(\theta^{(1,0)})^2 \{ 
h^{-\;a\underline{a}}_{++}v^{(0,1)}_a + \theta^{(0,1)\;\underline{b}} 
[ h^{(\underline{a}}_{++\;\underline{b})} +{1\over 2}\delta^
{\underline{a}}_{\underline{b}}(\partial_{--} h^{--}_{++} - 2 
h^{(ab)}_{++} v^{(0,1)}_av^{(0,-1)}_b) ]  \nonumber \\
&& + (\theta^{(0,1)})^2 ({1\over 2} 
t^{b\underline{a}}_{++-} -i\partial_{--}h^{-\;b\underline{a}}_{++} 
)v^{(0,-1)}_b \}~, \nonumber \\ 
H^{(2,0)}_R &=& i(\theta^{(1,0)})^2 \{ h_{++} + h^{(ab)}_{++}
v^{(0,1)}_av^{(0,-1)}_b - 
\theta^{(0,1)}_{\;\underline{a}}t_{++-}^{b\underline{a}}  \nonumber \\
&& - \;i(\theta^{(0,1)})^2 \partial_{--}h^{(ab)}_{++}
v^{(0,-1)}_av^{(0,-1)}_b \}~, 
\label{h20} 
\eea
\bea
\Lambda^{--} &=& \lambda^{--} - 2i \theta^{(0,1)}_{\;\underline{a}} 
\lambda^
{-\;a\underline{a}} v^{(0,-1)}_a + i(\theta^{(0,1)})^2 
\lambda^{(ab)}v^{(0,-1)}_av^{(0,-1)}_b \;,
\nonumber \\
\Lambda^{(0,1)\;\underline{a}} &=& \lambda^{-\;a\underline{a}}
v^{(0,1)}_a  + 
\theta^{(0,1)\;\underline{b}} [\lambda^{(\underline{a}}_{\;\;\underline{b})} + 
{1\over 2}\delta^{\underline{a}}_{\underline{b}} ( \partial_{--}
\lambda^{--} +2 \lambda^{(ab)}v^{(0,1)}_av^{(0,-1)}_b)] \nonumber \\
&&-(\theta^{(0,1)})^2 (i\partial_{--} \lambda^{-\;a\underline{a}} + 
{1\over 2}\beta^{a\underline{a}}_{-})v^{(0,-1)}_a \;, \nonumber \\
\Lambda_R &=& \lambda_R + \lambda^{(ab)}
v^{(0,-1)}_av^{(0,1)}_b                           
-\theta^{(0,1)}_{\;\underline{a}}\beta_{-}^{a\underline{a}}
v^{(0,-1)}_a  \nonumber \\
&& - \;i(\theta^{(0,1)})^2 \partial_{--}\lambda^{(ab)}
v^{(0,-1)}_av^{(0,-1)}_b \;, \label{l20} 
\eea
and $H^{(0,2)\;++}$, $H^{(1,2)\;\underline{i}}$, $H^{(0,2)}_L$, 
$\Lambda^{++}$, 
$\Lambda^{(1,0)\;\underline{i}}$, $\Lambda_L$ can be obtained from these 
expressions via the substitutions $+\leftrightarrow -$, 
$\theta^{(1,0)\;\underline{i}} \leftrightarrow \theta^{(0,1)\;
\underline{a}}$, 
$u \leftrightarrow v$, $i,\underline{i} \leftrightarrow a,\underline{a}$. 
In (\ref{h20}), (\ref{l20}) all the component fields and gauge parameters 
are functions of 
$z^{++}, z^{--}$ and we have explicitly indicated their $2D$ space-time 
indices. Note that in the chosen gauge the diagonal components of the 
world-sheet zweibein $h^{++}_{++}$, $h^{--}_{--}$ equal unity and 
the parameters of two independent Weyl rescalings of 
$\theta^{(1,0)\;\underline{i}}$, $\theta^{(0,1)\;\underline{a}}$ 
are fixed to be $\partial_{++}\lambda^{++}$, $\partial_{--}\lambda^{--}$, 
so the difference between the world and tangent indices of the involved 
fields actually disappears. Actually, we have used all the gauge symmetries
with  pure shifts in their transformation laws for gauging away the
corresponding  field components (rescalings are just of this kind). We ended
up only  with the transformations starting with $z$-derivatives of gauge
parameters.  

Looking at the above expressions we observe that the irreducible content of 
the original set of analytic vielbeins and connections includes only gauge 
fields: the two 
components of the world-sheet zweibein $h^{++}_{--}, h^{--}_{++}$, the
left and right gravitino components $h_{--}^{+\;i\underline{i}}, 
h_{++}^{-\;a\underline{a}}$, the left and right components of the 
$SO(4)_L\times U(1)_L$ and $SO(4)_R\times U(1)_R$ gauge connections   
$h_{--}^{(ij)}, h_{--}^{(\underline{i}\underline{j})}, h_{--}$ 
and $h_{++}^{(ab)}, h_{++}^{(\underline{a}\underline{b})}, h_{++}$, 
as well as the left and right components of the ``conformal gravitino'' 
$t_{--+}^{i\underline{i}}$, $t_{++-}^{a\underline{b}}$, with a total of 
(16 + 16) independent components.  
The remaining gauge freedom involves just the same number of gauge 
parameters, so locally all these gauge fields can be gauged away, though 
such a gauge is inadmissible globally (e.g., after coupling this 
multiplet to the $N=(4,4)$ string fields, the zweibein components should 
produce two Virasoro constraints). Therefore it is natural to call the 
obtained gauge multiplet, with no off-shell degrees of freedom, the ``
$N=(4,4),\; SO(4)\times U(1)$ Beltrami-Weyl (BW) multiplet''. 
We shall see later that 
it admits truncations to two different $N=(4,4),\; SU(2)$ ones. We will also 
show that the off-shell (8+8) ``minimal $N=4$, $2D$ SG multiplet'' 
\cite{{Gates1},{Gates2}} naturally comes out as the result of coupling
one of the  $N=(4,4)\;$, $SU(2)$ BW multiplets to one kind of twisted
$N=(4,4)$  multiplet treated as a compensator. 

Actually, in order to be able to construct manifestly invariant superfield 
couplings of $N=(4,4)$ BW  multiplets to $N=(4,4)$ matter, we 
need one more ingredient. This is an analytic density which 
should transform so as to cancel the transformation of the analytic 
superspace integration measure 
$\mu^{(-2, -2)}$. Indeed, 
as distinct from the flat superspace superconformal groups, the full local 
group (3.1) does not leave $\mu^{(-2,-2)}$ invariant:
\be \label{sgmeas}
\delta \mu^{(-2,-2)} = ((-1)^{P(\mu)}\partial_\mu\Lambda^\mu + 
\partial^{(-2,0)}\Lambda^{(2,0)} + \partial^{(0,-2)}
\Lambda^{(0,2)})\;\mu^{(-2,-2)} 
\equiv \tilde{\Lambda}\; \mu^{(-2,-2)}\;,
\ee
where $P(\mu)$ is 0 for bosonic and 1 for fermionic indices. 

Defining the objects   
\be  \label{gamma}
\Gamma^{(2,0)} = (-1)^{P(M)} \partial_M H^{(2,0)\;M}\;,\;\; 
\Gamma^{(0,2)} = (-1)^{P(M)} \partial_M H^{(0,2)\;M}\;, 
\ee
one finds them to transform as 
\begin{equation}
\delta \Gamma^{(2,0)} = \nabla^{(2,0)} \tilde{\Lambda}\;,\;\; 
\delta \Gamma^{(0,2)} = \nabla^{(0,2)} \tilde{\Lambda}
\end{equation}
and to satisfy, as a consequence of the constraints (\ref{constrH}), the 
condition 
\be  \label{constrgam}
\nabla^{(2,0)} \Gamma^{(0,2)} - \nabla^{(0,2)} \Gamma^{(2,0)} = 0\;. 
\ee
It is easy to show that (\ref{constrgam}) implies 
\be  \label{gamsigm}
\Gamma^{(2,0)} = \nabla^{(2,0)} \Sigma (\zeta,u,v)\;,\;\;
\Gamma^{(0,2)} = \nabla^{(0,2)} \Sigma (\zeta,u,v)\;.
\ee
Again, with making use of the constraints (\ref{constrH}), 
$\Sigma (\zeta, u,v)$ can be 
expressed in terms of the original BW multiplet (up to an unessential 
additive constant) \footnote{To the zeroth order in the
$\theta$'s and the first order in the fields, one has
$\Sigma = const + (h^{++}_{++} + h^{--}_{--}) +...$.} 
and shown to transform as
\be \label{trsigma}
\delta \Sigma = \tilde{\Lambda}~. 
\ee
Hence the quantity 
\be \label{compvol}
\Omega  \equiv e^{-\Sigma} \;,\;\;\delta \Omega  = -\tilde{\Lambda} \;
\Omega
\ee 
is the sought object, compensating for the non-invariance 
of the measure. In what follows we will need only the property 
\be \label{propgam}
(\,\nabla^{(2,0)} + \Gamma^{(2,0)}\,)\, \Omega = 0\;,\;\; 
(\,\nabla^{(0,2)} + \Gamma^{(0,2)}\,)\, \Omega = 0\;.
\ee
In particular, due to this property, one can still integrate by parts with 
respect to the {\it covariantized} harmonic derivatives. Indeed, for any 
analytic function $F(\zeta, u, v)$, the integral   
$$
\int \mu^{(-2,-2)} \Omega \nabla^{(2,0)}F(\zeta,u,v)~,   
$$
up to full {\it ordinary} derivatives, reduces to 
$$
- \int \mu^{(-2,-2)} (\nabla^{(2,0)} + \Gamma^{(2,0)})\Omega\; F(\zeta,u,v) 
= 0
$$
(the same is true for $\nabla^{(0,2)}$).
\setcounter{equation}{0}

\section{Various limits and truncations}
Inspecting the residual symmetry parameters \p{l20}, one observes that after 
constraining their $z$ dependence, in such a way that the left (right) 
parameters are functions solely of $z^{++} (z^{--})$, 
\bea
&& \partial_{--}\Lambda^{++} =  \partial_{++}\Lambda^{(1,0)\;\underline{i}} =
 \partial_{--}\Lambda_L = 0~, \nonumber \\
&& \partial_{++}\Lambda^{--} = \partial_{++}\Lambda^{(0,1)\;\underline{a}} 
=  \partial_{++}\Lambda_R = 0\;, \label{chirallim} 
\eea
they constitute the direct sum of 
two ``large''  $N=(4,4)$, $SO(4)\times U(1)$ superconformal algebras 
\cite{{Ademollo},{Belg2},{IKLev1},{IKLev2}}.
To see this, one should study the Lie
brackets of the transformations (3.1) into which these restricted
parameters  expanded in series in $z^{\pm\pm}$ are substituted. Then, e.g., for
the right branch,  one finds that the expansion of $\lambda^{--}(z^{--})$
produces a Virasoro  subsector, that of
$\lambda^{(\underline{a}\underline{b})}(z^{--}), \lambda^{(ab)}(z^{--})$
yields two affine $SU(2)$ subalgebras, and that of 
$\lambda^{-a\underline{b}}(z^{--}), \beta^{a\underline{b}}_-(z^{--})$ 
corresponds to the two types of SUSY generators present in this SCA, i.e. the
canonical generators (in particular, the $N=4$ Poincar\'e SUSY
and the special conformal SUSY generators) and the non-canonical ones. 
\footnote{Strictly speaking, such expansions define that part of 
$N=4$ SCA which is regular at the origin. Just such subalgebras of 
the left and right $N=4,\; SU(2)$ SCAs were gauged in the component 
approach of ref. \cite{schout}.} 
The affine $U(1)$ parameters contained in  $\lambda_R(z^{--})$ appear in the
closure of the canonical and non-canonical SUSY transformations (actually, 
the rigid $U(1)$ parameter $\lambda_R(z^{--})\vert_{z=0}$ never
appears in the closure on the superspace coordinates, but it
does appear when one considers 
the closure on the superfield $q^{(1,1)}$ with the transformation law 
\p{qutranloc}). It is also easy to check that these restricted 
superparameters coincide with those appearing in the realizations 
of these $N=4$ SCAs in the flat $SU(2)\times SU(2)$ harmonic superspace  
\cite{{ISu},{IS2}}.

Thus, we found that the original $N=(4,4)$ SG group \p{sgg}, \p{hdir20},
\p{hdir02}, \p{qutranloc}, \p{tranu1con} contains the direct sum of two 
$N=4, \; SO(4)\times U(1)$ SCAs as the essential 
invariance subalgebra of the residual gauge freedom associated 
with the superparameters \p{l20} (and their left counterparts). It should be 
stressed that it is an
invariance of the full nonlinear theory, not only of the linearized 
approximation \p{h20}. Indeed, it could be recovered from the general 
harmonic vielbein transformation laws \p{hdir20}, \p{hdir02}, 
\p{tranu1con}, as the maximal subgroup  preserving the flat limit 
\be \label{flat}
H^{(2,0)\;M} = H^{(0,2)\;M} = H^{(2,0)}_{L,R} = H^{(0,2)}_{L,R} = 0~.
\ee
Thus, the analytic superdiffeomorphism group of Sect. 3  
can be regarded as the local, gauged version of this maximal rigid $N=(4,4)$ 
superconformal group, with the BW multiplet defined by eq. \p{h20} 
(and by its left counterpart) as the corresponding gauge 
multiplet. Presumably, the latter can be alternatively recovered via direct 
gauging of this SCA following the procedure of ref. \cite{schout}. The 
$SU(2)\times SU(2)$ 
harmonic superspace approach allows one to relate it to the fundamental objects of 
the analytic superspace geometry, the analytic harmonic vielbeins 
$H^{(2,0)M}, H^{(0,2)M}$ and the analytic $U(1)$ connections 
$H^{(2,0)}_{L,R}, H^{(0,2)}_{L,R}$. 

Since $N=(4,4),\; SO(4)\times U(1)$ SCA contains as its infinite-dimensional 
subalgebras two $N=(4,4),\; SU(2)$ SCAs (SCA-I and SCA-II), 
it is natural to expect that its 
local extension also contains two smaller $N=(4,4)$ SG groups 
having these superconformal symmetries as the maximal ``rigid'' 
subgroups. They can 
naturally be called the $N=(4,4),\; SU(2)$ SG-I and SG-II groups. 
They should come out as appropriate truncations of \p{sgg}, \p{hdir20}, 
\p{hdir02}, \p{tranu1con} implemented through imposing certain 
constraints on the group parameters. The analytic harmonic 
vielbeins comprising the relevant 
shortened BW multiplets should then arise upon setting certain 
relations among the original analytic vielbeins, in a way covariant 
under the truncated SG group. 

One obvious truncation of the original group and vielbeins is as follows: 
\bea
&& \Lambda^{(2,0)} = \Lambda^{(0,2)} = \Lambda_L = 
\Lambda_R = 0~, \label{trun1} \\
&& H^{(4,0)} = H^{(0,4)} = H^{(2,2)} = \tilde{H}^{(2,2)} = H^{(2,0)}_{L,R} = 
H^{(0,2)}_{L,R} = 0~. \label{trunH}
\eea
The resulting group is the group of general analytic diffeomorphisms of 
the coordinates $\zeta^\mu$, with the inert harmonics 
\be
\delta \zeta^\mu = \Lambda^\mu(\zeta, u, v),\;\;\;\;\; 
\delta u = \delta v = 0~. 
\label{trun1SG}
\ee
The corresponding covariant harmonic derivatives read 
\be
\nabla^{(2,0)} = D^{(2,0)} + H^{(2,0)\;\mu}\partial_\mu~, \; 
\nabla^{(0,2)} = D^{(0,2)} + H^{(0,2)\;\mu}\partial_\mu~. \label{cov1SG}
\ee    
The transformation laws of these derivatives and vielbeins, as well as 
the constraints the latter should satisfy, directly follow from those 
given in the previous Section, after taking into account the constraints  
\p{trun1}, \p{trunH}. Note that the harmonic derivatives now are 
inert, 
$$
\delta \, \nabla^{(2,0)} = \delta \, \nabla^{(0,2)} = 0~, 
$$
and the integrability condition \p{consnab} becomes  
\be 
[\nabla^{(2,0)}, \nabla^{(0,2)}] = 0 \;\; \Rightarrow  
\label{cons1SG} 
\ee
\bea
&& \nabla^{(2,0)}H^{(0,2)\;++} - \nabla^{(0,2)}H^{(2,0)\;++} -2i
H^{(1,2)}\theta^{(1,0)} = 0~,  \nonumber \\
&& \nabla^{(2,0)}H^{(0,2)--} - \nabla^{(0,2)}H^{(2,0)--} +2i
H^{(2,1)}\theta^{(0,1)} = 0~, \nonumber \\
&& \nabla^{(2,0)}H^{(1,2)\;\underline{i}} - \nabla^{(0,2)}
H^{(3,0)\underline{i}} = 0~, \nonumber \\
&& \nabla^{(2,0)}H^{(0,3)\;\underline{a}} - \nabla^{(0,2)}
H^{(2,1)\;\underline{a}} = 0~. \label{constr1SG}
\eea

Comparing the above truncated transformations with those of 
the first rigid superconformal $N=4,\, SU(2)$ group (eqs. \p{scg2}, 
\p{tranharII}), one can suspect that the truncated SG group corresponds 
to gauging just this SCA-I. This is indeed the case. One can 
again choose the gauges 
\be
H^{(2,0)\;++} =  H^{(0,2)\;--} = H^{(0,3)\;\underline{a}} = 
H^{(3,0)\;\underline{i}}~, \label{gauges1SG}
\ee
as in \p{gauge1}, and repeat all the steps which led us 
to the irreducible field representation \p{h20} and the 
residual gauge freedom \p{l20}. For the truncated SG case 
we finally get, at the linearized level,  
\bea
H^{(2,0)\;--} &=& i(\theta^{(1,0)})^2 \{ h^{--}_{++} - 2i 
\theta^{(0,1)}_{\;\underline{a}} h^{-\;a\underline{a}}_{++} 
v^{(0,-1)}_a \}\;, \nonumber \\
H^{(2,1)\;\underline{a}} &=& i(\theta^{(1,0)})^2 \{ 
h^{-\;a\underline{a}}_{++}v^{(0,1)}_a + \theta^{(0,1)\;\underline{b}} 
[ h^{(\underline{a}}_{++\;\underline{b})} +{1\over 2}\delta^
{\underline{a}}_{\underline{b}}\;\partial_{--} h^{--}_{++}] \nonumber \\  
&& -\;i (\theta^{(0,1)})^2 \partial_{--}h^{-\;b\underline{a}}_{++} 
v^{(0,-1)}_b \} \;, \label{WZ1SG} 
\eea
\bea
\Lambda^{--} &=& \lambda^{--} - 2i \theta^{(0,1)}_{\;\underline{a}} 
\lambda^
{-\;a\underline{a}} v^{(0,-1)}_a~, \nonumber \\
\Lambda^{(0,1)\;\underline{a}} &=& \lambda^{-\;a\underline{a}}
v^{(0,1)}_a  + 
\theta^{(0,1)\;\underline{b}} [\lambda^{(\underline{a}}_{\;\;\underline{b})} + 
{1\over 2}\delta^{\underline{a}}_{\underline{b}}\;\partial_{--}
\lambda^{--}]
-i(\theta^{(0,1)})^2\partial_{--} \lambda^{-\;a\underline{a}}v^{(0,-1)}_a 
\label{Res1SG}
\eea
and analogous relations for the left vielbeins and parameters. 
We observe that the same can be obtained simply by setting 
$$
H^{(2,0)}_R = 0~, \;\; \Lambda_R = 0, \qquad 
H^{(0,2)}_L = 0~, \;\; \Lambda_L = 0 
$$ 
in the relations \p{h20}, \p{l20} (and their left counterparts). 
Thus we end up with the 
BW multiplet $h^{--}_{++}, h^{++}_{--}$, $h^{-\;a\underline{a}}_{++}, 
h^{+\;i\underline{i}}_{--}$, 
$h^{(\underline{a}\underline{b})}_{++}, 
h^{(\underline{i}\underline{k})}_{--}$ 
the field content of which basically coincides with that 
of the $N=(4,4),\quad SU(2)$ gauge multiplet found by Schoutens 
\cite{schout} (a slight difference 
comes from the fact that, on the way to this field representation, 
we have already gauge-fixed 
some local symmetries with pure shifts in the relevant gauge parameters, 
in particular, the local $2D$ Lorentz and scale 
invariances by setting $h^{++}_{++} = h^{--}_{--} = 1$). The 
residual gauge group has the parameters $\lambda^{--}, \lambda^{++}$ (local 
translations), $\lambda^{-\;a\underline{a}}, \lambda^{+\;i\underline{i}}$ 
(local supertranslations), $\lambda^{(\underline{a}\underline{b})}, 
\lambda^{(\underline{i}\underline{k})}$ (right and left $SU(2)$ groups). 
The number of these gauge invariances coincides with that of the gauge 
fields, so that the $N=(4,4),\; SU(2)$ BW multiplet (BW-I in what follows) 
contains no off-shell components like its parental 
$N=(4,4),\; SO(4)\times U(1)$ BW multiplet. 
Once again, the maximal subgroup of \p{trun1SG} preserving the flat 
limit
$$
H^{(2,0)\,\mu} = H^{(0,2)\,\mu} = 0
$$ 
is just the $N=(4,4),\; SU(2)$ SCA-I. It is singled out  
by imposing the light-cone chirality conditions on the parameters of  
the residual gauge group.

While specializing to the $N=(4,4),\; SU(2)$ SG-I group, 
we may retain the standard defining constraint for the twisted superfield 
$q^{(1,1)}$, 
\be \label{qudef1SG}
\nabla^{(2,0)}q^{(1,1)} = \nabla^{(0,2)}q^{(1,1)} = 0 
\ee 
(because of the commutativity property \p{cons1SG}), and the 
zero-weight scalar transformation rule 
\be \label{qutran1SG}
q^{(1,1)}{}'(\zeta{}', u, v) = q^{(1,1)}(\zeta, u, v)~. 
\ee
So, with respect to this SG-I group, $q^{(1,1)}$ is what is called TM-I in 
\cite{{Gates3},{gake}} because its physical bosonic fields $q^{ia}(z)$ are not 
affected by the local $SU(2)$ symmetries (on the contrary, 
the auxiliary fields $F^{\underline{i}\underline{a}}$ are transformed). 
Thus, the general rigidly supersymmetric  $q^{(1,1)}$ action \p{genaction} 
can be straightforwardly extended to the locally supersymmetric one 
\be  \label{locqact}
S_q^{I} = \int \mu^{(-2,-2)}\;\hat{\Omega}\;{\cal L}^{(2,2)}(q^{(1,1)M},u,v)~, 
\ee
where the density $\hat{\Omega}$ is still defined by eqs. 
\p{gamsigm}, \p{compvol}, with the truncation conditions  
\p{trunH} taken into account. In components and with the auxiliary fields 
eliminated, it gives the general locally supersymmetric $N=(4,4)$ 
sigma-model of ref. \cite{dWPVN} which is a modification of the sigma-model 
action of ref. \cite{PPVN} by torsion terms in the sector of the
physical bosons. For the rigid $q^{(1,1)}$ action \p{genaction}, 
the general torsionful off-shell component action was 
presented in \cite{ISu}. The action \p{locqact} yields a locally 
supersymmetric version of the latter. In Appendix we present, 
as an example, the component form of a very simple particular 
case of \p{locqact}.   

What about the second $N=(4,4),\; SU(2)$ SCA, with respect to which 
$q^{(1,1)}$ is TM-II? How to extract the relevant $N = (4,4)$ SG group 
from the original ``master'' SG group? It is easy to answer these
questions at the linearized level. The answer is prompted by the 
known realization of the $N=(4,4),\; SU(2)$ SCA-II in the 
$SU(2)\times SU(2)$ harmonic analytic superspace \cite{{ISu},{IS3}}. 
In order to have this SCA  
as the maximal symmetry after imposing the light-cone chiral 
constraints \p{chirallim}, one must seek for
restrictions on the residual gauge superparameters \p{l20} and their 
left counterparts such, that: i) the $U(1)$ 
parameters $\lambda_{L.R}$ are identified with $\partial_{\pm\pm}
\lambda^{\pm\pm}$; ii) the affine $SU(2)$ parameters 
$\lambda^{(\underline{a}\underline{b})}, 
\lambda^{(\underline{i}\underline{k})}$ are eliminated. 
The unique possibility to obey these requirements, still leaving the 
``true'' $SU(2)$ parameters $\lambda^{(ab)}, \lambda^{(ik)}$ unconstrained,   
is to impose the following relations:
\bea
\frac{\partial \Lambda^{(1.0)\;\underline{i}}}
{\partial {\theta^{(1,0)\;\underline{k}}}} =  
\delta^{\underline{i}}_{\underline{k}}\;
(\Lambda_L + \partial_{++}\Lambda^{++})~, \quad
\frac{\partial \Lambda^{(0,1)\;\underline{a}}}
{\partial {\theta^{(0,1)\;\underline{b}}}} = 
\delta^{\underline{a}}_{\underline{b}}\;(\Lambda_R + 
\partial_{--}\Lambda^{--})~,  \label{trun2}     
\eea
whence 
\bea
&&\lambda^{(\underline{a}\underline{b})} = 
\lambda^{(\underline{i}\underline{k})} = 0~, \quad
\lambda_L = -{1\over 2}\,\partial_{++}\lambda^{++}~, \;
\lambda_R = -{1\over 2}\,\partial_{--}\lambda^{--}~, \nonumber \\ 
&& \beta_+^{i\underline{k}}= 
-2i\partial_{++}\lambda^{+\;i\underline{k}}~, \;\;\;
\beta_-^{a\underline{b}}= 
-2i\partial_{--}\lambda^{-\;a\underline{b}}~. \label{trun2par} 
\eea
It is easy to explicitly check that the superparameters \p{l20} 
(and their left counterparts) restricted in this way indeed span 
the sought $N=(4,4),\, SU(2)$ SCA-II after imposing the chirality 
conditions \p{chirallim}. 
Then, at the linearized level, it is a consistent truncation 
to set equal to zero those combinations of the analytic 
vielbeins, which are not shifted under the subgroup 
singled out by eqs.\p{trun2}:
\bea
\frac{\partial H^{(1.2)\;\underline{i}}}
{\partial {\theta^{(1,0)\;\underline{k}}}} =  
\delta^{\underline{i}}_{\underline{k}}\;
(\partial_{++}H^{(0,2)\;++}- H^{(0,2)}_L)~, \;
\frac{\partial H^{(2,1)\;\underline{a}}}
{\partial {\theta^{(0,1)\;\underline{b}}}} = 
\delta^{\underline{a}}_{\underline{b}}\;(\partial_{--}H^{(2,0)\;--} - 
H^{(2,0)}_R)~.  \label{trun2H}     
\eea
These relations amount to the following linearized constraints on the gauge 
fields:
\bea 
&& h^{(\underline{a}\underline{b})}_{++} =
h^{(\underline{i}\underline{k})}_{--} 
= 0~, \quad h_{--} = {1\over 2}\,\partial_{++}h^{++}_{--}~, \; 
h_{++} = {1\over 2}\,\partial_{--}h^{--}_{++}~, \nonumber \\
&& t^{i\underline{i}}_{--+} = 2i\partial_{++}h^{+\;i\underline{i}}_{--}~, \;\;
t^{a\underline{a}}_{++-} = 2i\partial_{--}h^{-\;a\underline{a}}_{++}~. 
\label{trun2h} 
\eea
They leave us with the representation $h^{--}_{++}, h^{++}_{--}$, 
$h^{+\;i\underline{k}}_{--}$, $h^{-\;a\underline{b}}_{++}$, 
$h^{(ik)}_{--}, h^{(ab)}_{++}$, which is again 
a $N=(4,4),\; SU(2)$ BW multiplet, but with another 
chiral pair of $SU(2)$ gauge fields, compared to \p{WZ1SG}.
We call it the $N=(4,4),\; SU(2)$ BW-II multiplet.
For completeness, we explicitly quote here the counterparts of 
\p{WZ1SG}, \p{Res1SG} for the considered case
\bea
H^{(2,0)\;--} &=& i(\theta^{(1,0)})^2 \{ h^{--}_{++} - 2i 
\theta^{(0,1)}_{\;\underline{a}} h^{-\;a\underline{a}}_{++} 
v^{(0,-1)}_a -i 
(\theta^{(0,1)})^2 h^{(ab)}_{++} v^{(0,-1)}_av^{(0,-1)}_b \}\;, \nonumber \\
H^{(2,1)\;\underline{a}} &=& i(\theta^{(1,0)})^2 \{ 
h^{-\;a\underline{a}}_{++}v^{(0,1)}_a +{1\over 2}\,
 \theta^{(0,1)\;\underline{a}}(\partial_{--} h^{--}_{++} - 2 
h^{(ab)}_{++} v^{(0,1)}_av^{(0,-1)}_b)\}~, \nonumber \\
H^{(2,0)}_R &=& i(\theta^{(1,0)})^2 \{ {1\over 2}\,\partial_{--}h^{--}_{++} 
+ h^{(ab)}_{++}
v^{(0,1)}_av^{(0,-1)}_b - 
2i\,\theta^{(0,1)}_{\;\underline{a}}\partial_{--}
h^{-b\underline{a}}_{++} \nonumber \\
&& - \;i(\theta^{(0,1)})^2 \partial_{--}h^{(ab)}_{++}
v^{(0,-1)}_av^{(0,-1)}_b \}~, 
\label{h2SG} 
\eea
\bea
\Lambda^{--} &=& \lambda^{--} - 2i \theta^{(0,1)}_{\;\underline{a}} 
\lambda^
{-\;a\underline{a}}\, v^{(0,-1)}_a + i(\theta^{(0,1)})^2 
\lambda^{(ab)}v^{(0,-1)}_av^{(0,-1)}_b~, 
\nonumber \\
\Lambda^{(0,1)\;\underline{a}} &=& \lambda^{-\;a\underline{a}}
\,v^{(0,1)}_a  + 
{1\over 2}\,\theta^{(0,1)\;\underline{a}}( \partial_{--}
\lambda^{--} +2 \lambda^{(ab)}v^{(0,1)}_av^{(0,-1)}_b)~, \nonumber \\
\Lambda_R &=& -{1\over 2}\,\partial_{--}\lambda^{--} + \lambda^{(ab)}
v^{(0,-1)}_av^{(0,1)}_b +                          
 2i\,\theta^{(0,1)}_{\;\underline{a}}\partial_{--}\lambda^{-\;a\underline{a}}
\,v^{(0,-1)}_a \nonumber \\
&& - \;i(\theta^{(0,1)})^2 \partial_{--}\lambda^{(ab)}
v^{(0,-1)}_av^{(0,-1)}_b \;. \label{l2SG} 
\eea
The left objects are obtained via the same substitutions as 
in the previous cases. 

For the time being, we do not know how to go beyond the linearized level 
in this important case. It seems that it is more fruitful to descend 
to the above shortened versions of the BW multiplets (and further to 
the Poincar\'e SG), using a more convenient approach based on  
the concept of superconformal compensation. 

\setcounter{equation}{0}

\section{Superconformal matter couplings}\label{supmat} 
The basic idea of the compensation approach (see, e.g., \cite{book}) 
is to start from the pure superconformal SG and then to couple to it, 
in a superconformally covariant way, 
appropriate matter multiplets with inhomogeneous (Goldstone type) 
transformation laws with respect to certain (super)conformal symmetries. 
Then, by properly fixing gauges (normally, in such a way that all 
inhomogeneously transforming components are fully gauged away), one gets as 
a net result the theory with a smaller number of local symmetries and
supersymmetries, i.e. a sort of Poincar\'e SG. The auxiliary fields of 
the compensating superfield become in this gauge auxiliary fields of 
the relevant Poincar\'e SG gauge multiplet. If, from the beginning, 
a few matter superfields coupled to a given conformal SG are included, 
being one of them a compensator, we end up with the theory of the remaining 
matter multiplets in a Poincar\'e SG background. In this way, one can derive 
various Poincar\'e-type supergravities (with all, or a part of, the 
original conformal symmetries compensated for), different off-shell SG 
multiplets (depending on the choice of compensator), etc.     
 
We believe that the $N=(4,4),\; SO(4)\times U(1)$ SG group defined in Sect. 
3 is the maximal, ``master'' $N=(4,4),\; 2D$ conformal SG group. 
Then, the relevant gauge multiplet, $N=(4,4),\; SO(4)\times U(1)$ BW 
multiplet, is the ``master'' multiplet from which all other known 
$N=(4,4)$ SG multiplets should follow by the appropriate 
compensating procedure. To list all possibilities, we need to 
know all possible superconformal rigid off-shell matter multiplets 
which can be defined in $SU(2)\times SU(2)$ harmonic superspace, 
their off-shell 
actions, and the locally superconformal extensions of the latter. As
it was already noticed earlier, not all known types of twisted
superfields (and their variant representations) admit a simple
formulation in $SU(2)\times SU(2)$ analytic harmonic superspace
\cite{IS2}. In what follows, we shall deal with 
the superconformal off-shell matter multiplets which admit a description 
in terms of analytic $SU(2)\times SU(2)$ harmonic superfields and 
which were reviewed in Sect. 2. These are the nonlinear multiplets 
$N^{(2,0)}, N^{(0,2)}$, $G^{(2,0)}, G^{(0,2)}$ and the twisted 
chiral multiplets $q^{(1,1)}$ which can be either TM-I or TM-II,   
depending on the superconformal $N=(4,4),\; SU(2)$ group with respect
to which one studies their transformation properties.  
We shall show that 
some of these superfields can be used to compensate the ``master'' 
$N=(4,4)$ conformal SG group down to its $N=(4,4),\; SU(2)$ 
subgroups and, further, to the Poincar\'e SG groups, including 
the group of minimal off-shell SG of refs. \cite{{Gates1},{Gates2}}. 

We start with a local extension of the set $N^{(2,0)}, N^{(0,2)}$. The 
rigid superconformal transformation laws of this multiplet \p{transfQ} 
naturally generalize to the whole ``master''$N=(4,4)$ SG group as  
\be
\delta \,N^{(2,0)} = \Lambda^{(2,0)}~, \;\;\;
\delta \,N^{(0,2)} = \Lambda^{(0,2)}~, \label{localN}
\ee 
where the transformation parameters are now the general analytic
superfunctions introduced in (3.1). The defining constraints 
\p{flatQconstr} are covariantized as follows: 
\bea
&&(a)\quad \nabla^{(2,0)}N^{(2,0)} + N^{(2,0)}N^{(2,0)} = H^{(4,0)}~, \; 
\nabla^{(0,2)}N^{(0,2)} + N^{(0,2)}N^{(0,2)} = H^{(0,4)}~, \nonumber \\
&&(b)\quad \nabla^{(2,0)}N^{(0,2)} - \nabla^{(0,2)}N^{(2,0)} = H^{(2,2)} - 
\tilde{H}^{(2,2)} \label{constrNcurv}
\eea
(for a similar covariantization of the standard nonlinear multiplet 
in the conventional harmonic superspace, see \cite{{GIOS2},{IS3}}). 
It is obvious that the
$N$-multiplet can be used to fully compensate all gauge invariances 
contained in $\Lambda^{(2,0)}, \Lambda^{(0,2)}$, including two chiral 
$SU(2)$ symmetries acting on the harmonic variables. One can achieve
this purpose, choosing the gauge 
\be
N^{(2,0)} = N^{(0,2)} = 0 \quad \Rightarrow \quad (a)\;H^{(0,4)} = 
H^{(4,0)} = 0~, 
\;\;(b)\; H^{(2,2)} - \tilde{H}^{(2,2)} = 0~. \label{gaugeN}
\ee              

Prior to any gauge-fixing, it is instructive to fully 
elaborate on the corollaries of the constraints 
\p{constrNcurv}. For the quantities
\be \label{defQ}
Q^{(2,0)} \equiv N^{(2,0)} - H^{(2,0)}_L - H^{(2,0)}_R~, \quad 
Q^{(0,2)} \equiv N^{(0,2)} - H^{(0,2)}_L - H^{(0,2)}_R~,
\ee   
eq. (\ref{constrNcurv}$b$), combined with eqs. \p{constrh22} implies 
the following constraint:
\be \label{constrQQ}
\nabla^{(2,0)}Q^{(0,2)} - 
\nabla^{(0,2)}Q^{(2,0)}= 0 \quad \Rightarrow \quad Q^{(2,0)} = 
\nabla^{(2,0)}\Phi~, \;Q^{(0,2)} =  \nabla^{(0,2)}\Phi~, 
\ee
where $\Phi = \Phi(\zeta,u,v)$ is a new analytic compensating superfield. 
Recalling the transformation properties \p{tranu1con}, \p{localN}, we see 
that 
\be 
\delta\, Q^{(2,0)} = \nabla^{(2,0)}(\Lambda_L + \Lambda_R)~, \;\; 
\delta\, Q^{(0,2)} = \nabla^{(0,2)}(\Lambda_L + \Lambda_R) \;\; \Rightarrow   
\; \delta\, \Phi = \Lambda_L + \Lambda_R~. \label{tranPhi} 
\ee
Hence, the newly introduced analytic object $\Phi$ can be fully gauged 
away using the analytic gauge parameter
$\Lambda_L+\Lambda_R$
\be \label{gaugePhi}
\Phi = 0 \quad \Rightarrow \quad \Lambda_L = - \Lambda_R \equiv \Lambda~.
\ee
As a corollary of this choice, the following relations occur:
\be \label{NHH}
Q^{(2,0)}= Q^{(0,2)} = 0 \quad \Rightarrow \quad 
N^{(2,0)} = H^{(2,0)}_L + H^{(2,0)}_R ~, \;\; 
N^{(0,2)} = H^{(0,2)}_L + H^{(0,2)}_R~. 
\ee

At this stage, it is time to fix the gauge freedom associated 
with the superparameters $\Lambda^{(2,0)}, \Lambda^{(0,2)}$, by imposing 
the gauge \p{gaugeN}. As a result of this gauge choice, the 
original ``master'' $N=(4,4)$ SG group (3.1) proves to be compensated 
just down to its $N=(4,4),\; SU(2)$ SG-I subgroup \p{trun1SG}. 
Eqs. \p{NHH}, in this gauge, imply 
\bea 
&& H^{(2,0)}_L = - H^{(2,0)}_R \equiv H^{(2,0)}~, \quad 
H^{(0,2)}_L = - H^{(0,2)}_R \equiv H^{(0,2)}~, \label{restrH} \\
&& \delta\,H^{(2,0)} = 
\nabla^{(2,0)}\Lambda~, \;
\delta\,H^{(0,2)} = 
\nabla^{(0,2)}\Lambda~. \label{tranHnew}
\eea 
As a consequence of these relations and the gauge choice \p{gaugePhi}, 
the transformation law \p{qutranloc} of the twisted multiplet $q^{(1,1)}$ 
in the ``master'' SG group, as well as its defining constraints 
\p{curqucons}, are reduced to those covariant under the
$N=(4,4),\; SU(2)$ SG 
group, i.e. \p{qutran1SG} and \p{qudef1SG}. Nevertheless, the 
resulting theory is not yet identical to what we have got after 
truncation in Sect. 4. Indeed, the gauge-fixed 
covariant derivatives $\nabla^{(2,0)}, \nabla^{(0,2)}$ differ from those 
defined by eq. \p{cov1SG}  
\bea
&&\nabla^{(2,0)} = D^{(2,0)} + H^{(2,0)\;\mu}\partial_\mu + 
H^{(2,2)}\partial^{(0,-2)}~, \nonumber \\ 
&&\nabla^{(0,2)} = D^{(0,2)} + H^{(0,2)\;\mu}\partial_\mu +
H^{(2,2)}\partial^{(-2,0)}~,\label{modnabl} \\ 
&& H^{(2,2)} = \nabla^{(2,0)}H^{(0,2)} -
\nabla^{(0,2)}H^{(2,0)}~.
\label{defHnew} 
\eea
Though $H^{(2,2)}$ as well as $H^{(2,0)}, \; H^{(0,2)}$ transform as 
scalars under the remaining $N=(4,4),\;SU(2)$ SG-I group, the harmonic 
partial derivatives $\partial^{(0,-2)}, \partial^{(-2,0)}$ are not 
covariant, due to the presence 
of a non-trivial $u,v$ dependence in the group parameters in \p{trun1SG}. As 
a result, $H^{(2,2)}$ appears in the transformation laws 
of the vielbeins $H^{(2,0)\;\mu}, H^{(0,2)\;\mu}$. The constraints 
\p{constrH} also do not go into the set \p{constr1SG}, due to the presence 
of $H^{(2,2)}$. For this object, the original constraints \p{constrH}  
imply the following ones (recall that $H^{(4,0)}=H^{(0,4)} = 0$ 
in the gauge \p{gaugeN}): 
\be  \label{residconstr}
\nabla^{(2,0)}H^{(2,2)} = \nabla^{(0,2)}H^{(2,2)} = 0~.
\ee  

This peculiarity comes out only at the nonlinear level.  
The linearized analysis goes as before and shows that $H^{(2,0)\;\mu}, 
H^{(0,2)\;\mu}$, in the present case, carry the same set of fields 
forming the $N=(4,4),\;SU(2)$ BW-I multiplet. In other words, after fixing  
appropriate conformal gauges in the locally superconformal system of 
the original ``master'' BW multiplet and the compensator multiplet 
$N^{(2,0)}, N^{(0,2)}$, we are left with a smaller $N=(4,4),\; SU(2)$ 
BW-I multiplet and an extra off-shell multiplet. The latter is carried 
by the superfields $H^{(2,0)}, H^{(0,2)}$ which exhibit the gauge 
freedom \p{tranHnew} with an extra analytic gauge parameter
$\Lambda(\zeta,u,v)$ and satisfy the constraints \p{residconstr}.
This extended representation is not fully reducible, in the sense 
that the additional gauge superfields $H^{(2,0)}, H^{(0,2)}$ are scalars 
with respect to the conformal $N=(4,4),\;SU(2)$ SG-I group \p{trun1SG} 
while the SG-I transformation laws of the analytic vielbeins $H^{(2,0)\;\mu}, 
H^{(0,2)\;\mu}$ include these extra superfields.    

Thus, we have found the previously  unknown off-shell $N=(4,4)$ 
SG gauge multiplet. In the WZ gauge and at the linearized level, 
its part coming from the analytic 
vielbeins is the same BW-I gauge fields representation which was described 
in Sect. 4 and which has no off-shell degrees of freedom (the linearized 
structure \p{WZ1SG} in this case is slightly modified by the fields from 
$H^{(2,0)}, H^{(0,2)}$, because of the presence of $H^{(2,2)}$ in the 
constraints on $H^{(0,2)\;\mu}, H^{(2,0)\;\mu}$). To examine the off-shell 
content of $H^{(2,0)}, H^{(0,2)}$, we have chosen an appropriate 
WZ gauge with respect to the parameter $\Lambda(\zeta,u,v)$, 
so as to kill as much component fields in the $\theta $, $u,v$ expansions 
of these superfields as possible, and inserted  the result into the 
linearized form of the constraints \p{residconstr}. Solving the latter (it 
does not put any field on shell), we have eventually found $(32 + 32)$ 
independent off-shell components listed 
below (the numerals in the parentheses on the right to the
fields denote the ``engineering'' dimension and the number
of independent real components, respectively):
\bea
&&\underline{\mbox{bosons}}: \;\; (h_{++}, h_{--})\; (1,1)~, \;\;
l^{(ab)}_{++}\;(1,3)~, \;\;l^{(ik)}_{--}\;(1,3)~, \;\; 
l^{ia}_{\underline{i}\underline{a}}\;(1,16)~, \;\;l^{(ik)(ab)}\;(0,9)~, 
\nonumber \\
&&\underline{\mbox{fermions}}: \;\; l^b_{\underline{a}\;+}\;(3/2,4)~, \;\;
l^i_{\underline{k}\;-}\;(3/2,4)~, \;\; l^{(ik)\;a}_{\underline{b}}\;(1/2,12)~, 
\;\; l^{(ab)\;i}_{\underline{k}}\;(1/2,12)~. \label{bosfer}
\eea
The fields $h_{\pm\pm}$ are gauge fields for a $U(1)$ with the gauge 
parameter $\lambda(z)$ which is the first component in  
$\Lambda(\zeta,u,v)$. This $U(1)$ is the only residual gauge symmetry 
of the given WZ gauge. The fields $l^{(ab)}_{++}~,\; l^{(ik)}_{--}$ 
and $l^b_{\underline{a}\;+}~,\;l^i_{\underline{k}\;+}$ are ``former'' 
gauge fields for the symmetries with the parameters $\lambda^{(ab)}, 
\lambda^{(ik)}$ and $\beta^b_{\underline{a}\;-},\;
\beta^i_{\underline{k}\;+}$ in the ``master'' BW multiplet 
(eqs. \p{h20}, \p{l20} and their left counterparts). Now these 
local symmetries have been entirely compensated by the appropriate 
compensating fields from $N^{(2,0)}, N^{(0,2)}$. Note that the 
residual gauge group $U(1)$ is the diagonal in the product of 
two chiral gauge $U(1)$ groups realized on the ``master'' BW multiplet; the 
rest of these $U(1)$ symmetries has been compensated by 
a dimension-0, $SO(4)$ singlet field present 
in $N^{(2,0)}, N^{(0,2)}$ \cite{IS3} (this is just the first component 
of the compensator $\Phi$ introduced in \p{constrQQ}). The biggest flat limit 
symmetry of the extended gauge multiplet ($N=(4,4),\;SU(2)$ BW-I  
together with  \p{bosfer}) is $N=(4,4),\;SU(2)$ SCA-I augmented with 
an extra rigid $U(1)$ symmetry. Note that in the 
matter couplings we shall discuss in this Section, 
the superfields $H^{(2,0)}, H^{(0,2)}$ always 
appear only through their analytic superfield strength $H^{(2,2)}$ 
containing, in particular, the field strength of the $U(1)$ 
gauge field $h_{\pm\pm}$. In other words, the residual local $U(1)$ 
group is hidden, and for the time being we do not see in which 
situations it could become active.

A comment is to the point here. In principle, we could completely 
eliminate the extra multiplet by treating it as pure gauge. 
This possibility corresponds to adding additional constraints 
to the set \p{constrNcurv}
\be
H^{(2,2)} = \nabla^{(2,0)}N^{(0,2)}~, \quad 
\tilde{H}^{(2,2)} = \nabla^{(0,2)}N^{(2,0)}~. \label{modconstr}
\ee
These constraints are manifestly covariant and compatible  
both with \p{constrNcurv} and \p{constrH}.The same reasoning which led 
us to eqs. \p{constrQQ}, \p{restrH} implies that in the gauge \p{gaugeN} 
the pairs $H^{(2,0)}_L, H^{(0,2)}_L$ and $H^{(2,0)}_R, H^{(0,2)}_R$ 
become pure gauge, with respect to the $U(1)$ gauge groups with parameters 
$\Lambda_L$ and $\Lambda_R$. Hence, they can be gauged away, 
fully compensating this gauge freedom. As the result, the
$N=(4,4),\,SU(2)$ 
SG-I group and the BW-I multiplet are finally reproduced. 
A deviation from the standard compensation point of view is that, 
after imposing  \p{modconstr}, the compensators $N^{(2,0)}, N^{(0,2)}$ 
cease to have a flat off-shell limit (when all vielbeins 
are put equal to zero): the resulting modified set of constraints proves  
to be too restrictive, it puts these superfield on shell \cite{IS3}. 
On the other hand, one can view the relations \p{constrNcurv} and 
\p{modconstr} merely as the covariant definition of 
particular harmonic vielbeins $H^{(0,4)}, H^{(4,0)}, H^{(2,2)}, 
\tilde{H}^{(2,2)}$, such that it provides  
a covariant way to make some gauge fields in the ``master'' BW 
multiplet purely longitudinal and, so, globally removable 
by fixing appropriate gauges. Indeed, from the standpoint of 
the linearized WZ representation \p{h20} for the ``master'' BW multiplet, 
these relations mean that all gauge fields except those comprising 
the $N=(4,4),\;SU(2)$ BW-I multiplet are postulated to be pure gauge. 

Let us now turn to the issue of constructing matter actions invariant 
under the ``master'' conformal SG group. 

We start by seeking for the appropriate generalization of the 
$N$-action \p{lagrN}. Somewhat surprisingly, it cannot 
be straightforwardly promoted to an invariant of the
local superconformal group. The best we have reached,
in our attempts to covariantize \p{lagrN}, is the action 
\be \label{trialact}
S^{loc}_N  = -\int \mu^{(-2,-2)}\, \Omega \left(Q^{(2,0)}Q^{(0,2)} + 
2Q^{(2,0)}H^{(0.2)}_L + 2Q^{(0,2)}H^{(2,0)}_R + 
2H^{(0,2)}_LH^{(2,0)}_R\right), \label{noninv}
\ee  
where $Q^{(2,0)}, Q^{(0,2)}$ are defined in \p{defQ}. 
It is shifted, up to surface terms, by the expression 
\be \label{shift}
\delta\,S^{loc}_N = - 2 \,\int \mu^{(-2,-2)}\,\Omega\,\left(H^{(0,2)}_L
\nabla^{(2,0)}\Lambda_L + H^{(2,0)}_R\nabla^{(0,2)}\Lambda_R \right)~, 
\ee  
which cannot be further cancelled in any way. 

On the other hand, it is possible to construct invariant actions 
for the second type of superconformally invariant (32+32) nonlinear 
multiplet defined by \p{rigG1}, \p{flGconstr}. The constraints \p{flGconstr} 
admit a direct covariantization
\bea
&& (\nabla^{(2,0)} + 2N^{(2,0)})G^{(2,0)} + 
\alpha \,G^{(2,0)}G^{(2,0)} =0~, \nonumber \\
&& (\nabla^{(0,2)} + 2N^{(0,2)})G^{(0,2)} + 
\alpha\,G^{(0,2)}G^{(0,2)} =0~, \nonumber \\
&& \nabla^{(2,0)}G^{(0,2)} - \nabla^{(0,2)}G^{(2,0)} = 0~. 
\label{curGconstr}      
\eea
Indeed, it is easy to check their covariance under (3.1) 
provided that the superfields $G$ transform as scalars: 
$\delta\,G^{(2,0)} = \delta\, G^{(0,2)} = 0$. Then the simplest
manifestly invariant action of $G^{(2,0)}, G^{(0,2)}$ in the background 
of the $N=(4,4)$ ``master'' conformal SG fields and compensators 
$N^{(2,0)}, N^{(0,2)}$ is given by 
\be 
S^{loc}_G = - \int \mu^{(-2,-2)}\,\Omega \, G^{(2,0)}G^{(0,2)}~. \label{GG} 
\ee

Another possibility to construct an invariant off-shell action for the 
pair of compensators $N^{(2,0)}, N^{(0,2)}$ is to take as the relevant 
Lagrangian density the constraints \p{constrNcurv} with the appropriate 
analytic Lagrange multipliers $\omega^{(-2,2)}$, $\omega^{(2,-2)}$, 
$\omega $. Just an action of this kind describes the standard nonlinear 
multiplet coupled to a conformal $N=2, \;4D$ SG in the conventional harmonic 
superspace \cite{{GIKOS},{GIOS2}}. Its $SU(2)\times SU(2)$ analogue would 
also have no propagating degrees 
of freedom and, before varying with respect to Lagrange multipliers, contain
an infinite number of auxiliary fields. This possibility  requires a thorough
analysis and we postpone discussing it to the future.   

It is worth noting that there are no problems with extending 
the flat superspace actions of $N^{(2,0)}, N^{(0,2)}$ and 
$G^{(2,0)}, G^{(0,2)}$ to invariants of the $N=(4,4)$ SG-I group. 
The relevant constraints are obtained from the flat ones 
\p{flatQconstr}, \p{flGconstr} by the replacements 
$D^{(2,0)}, D^{(0,2)} \;\rightarrow \; 
\nabla^{(2,0)}, \nabla^{(0,2)}$, where $\nabla^{(2,0)}, \nabla^{(0,2)}$
are given by eqs. \p{cov1SG}, and the locally supersymmetric actions 
are obtained via the replacement $\mu^{(-2,-2)} \;\rightarrow\; 
\mu^{(-2,-2)}\,\hat{\Omega}$ in the flat superspace ones. 

Let us now switch over to the twisted multiplets. We already constructed in 
Sect. 4 a locally supersymmetric $q^{(1,1)}$ action \p{locqact} invariant 
under the
$N=(4,4),\;SU(2)$ SG-I group. An important question is how to construct 
the $q^{(1,1)}$ actions invariant under the full ``master'' $N=(4,4)$ 
group (3.1). The main difficulty here is related to the non-trivial 
transformation law \p{qutranloc} of $q^{(1,1)}$ in this group. 
 
The simplest way to construct such a coupling is to consider $q^{(1,1)}$ 
together with the compensators $N^{(2,0)}, N^{(0,2)}$. In this case, 
due to the existence of the analytic scalar compensator $\Phi$ which 
is shifted by the sum $\Lambda_L + \Lambda_R$ (eq. \p{constrQQ}), one can 
redefine any $q^{(1,1)}$ with the transformation law \p{qutranloc} in 
such a way that it will transform as a scalar under the ``master'' SG 
group 
\be
q^{(1,1)}(\zeta,u,v) \quad \Rightarrow \quad \tilde{q}^{(1,1)} = 
e^\Phi\;q^{(1,1)}~, \;\; \tilde{q}^{(1,1)}{}' (\zeta', u', v') = 
\tilde{q}^{(1,1)}(z,u,v)~.
\ee
The constraints \p{curqucons} become 
\be
(\nabla^{(2,0)} + N^{(2,0)})\tilde{q}^{(1,1)} = 0~, \quad 
(\nabla^{(0,2)} + N^{(0,2)})\tilde{q}^{(1,1)} = 0~.  
\ee     
Their covariance is evident. The general invariant action is similar to 
\p{locqact}
\be
\tilde{S}^I_q = 
\int \mu^{(-2,-2)}\;\Omega\;{\cal L}^{(2,2)}(\tilde{q}^{(1,1)}, \tilde{u}, 
\tilde{v})~, \label{tildeact}
\ee
where \footnote{Within the conventional harmonic superspace,
the necessity of analogous 
redefinitions of the harmonics explicitly appearing in the action of 
hypermultiplets coupled to conformal $N=2$, $4D$ SG was firstly 
shown in \cite{GIOS2}.} 
\bea
&& \tilde{u}^{(1,0)} = u^{(1,0)} - N^{(2,0)}u^{(-1,0)}~, \quad 
\tilde{u}^{(-1,0)} = u^{(-1,0)}~, \nonumber \\ 
&& \tilde{v}^{(0,1)} = v^{(0,1)} - N^{(0,2)}v^{(0, -1)}~, \quad
\tilde{v}^{(-1,0)} = v^{(0,-1)}~.
\eea
In the gauge \p{gaugeN} the action \p{tildeact} coincides with \p{locqact} 
modulo 
a modification of both the covariant harmonic derivatives and the 
constraints on the analytic vielbeins due to the presence of the 
$U(1)$ gauge multiplet $H^{(2,0)}, H^{(0,2)}$. The effect of 
this modification is two-fold: first, the constraints defining 
$q^{(1,1)}$ are obscured by this extra multiplet and, second, 
the density $\Omega$ differs from $\tilde{\Omega}$ 
in \p{locqact} owing to the presence of the extra multiplet in 
the constraints for the analytic vielbeins. 
It would be interesting 
to see what is the precise impact of this modification on the component 
sigma-model action as compared to the action \p{locqact} which includes the 
$N=(4,4),\,SU(2)$ BW-I multiplet without any additional SG fields.  
   
Since there exists the unique $N=(4,4)$ WZW $q^{(1,1)}$ action \p{wzwact} 
invariant under the full rigid $N=(4,4),\;SO(4)\times U(1)$ superconformal 
symmetry, it is natural to seek for its direct coupling to the ``master'' 
$N=(4,4)$ BW multiplet without adding any extra compensators. If such a 
coupling can be set up, $q^{(1,1)}$ can be regarded, like $N^{(2,0)}, 
N^{(0,2)}$, as a compensator extending the master 
BW multiplet to some SG multiplet with a smaller number of gauge symmetries 
and gauge fields. The corresponding SG group should be some 
subgroup of the master $N=(4,4)$ SG group. Indeed, 
the shifted superfield $\hat{q}^{(1,1)}$ defined 
in \p{defc11} 
transforms inhomogeneously under \p{sgg}, \p{qutranloc}
\be \label{tranhat}
\delta\,\hat{q}^{(1,1)} = (\Lambda_L + \Lambda_R)(\hat{q}^{(1,1)} + 
c^{(1,1)}) - \Lambda^{(2,0)}c^{(-1,1)} - \Lambda^{(0,2)}c^{(1,-1)}~, 
\ee      
and hence it can be employed as a compensator.

Unfortunately, we do not have yet any general recipe how to construct 
such a locally supersymmetric  extension of \p{wzwact}. The main 
difficulty stems from the fact that the analytic superfield density in 
\p{wzwact} is not a tensor: it is shifted by full harmonic derivatives 
under the rigid superconformal $SO(4)\times U(1)$ transformation. 
The most straightforward approach is to restore the full action 
order by order in the SG superfields, and this is what we shall undertake. 

First, we make the replacement 
$$
\mu^{(-2,-2)} \quad \Rightarrow \quad \mu^{(-2,-2)}\;\Omega 
$$
in \p{wzwact} in order to be able to integrate by parts with respect 
to $\nabla^{(2,0)}$, $\nabla^{(0,2)}$ (recall the discussion at the end 
of Sect. 3). We do not fix beforehand any gauges including \p{basgauge}. 
Thus we represent the sought $S_{wzw}^{loc}$ as a series in powers 
of the SG superfields 
\be \label{iterat}
S^{loc}_{wzw} = S_{(0)} + S_{(1)} + S_{(2)} + ... = 
-{1\over 4\gamma^2}\;\int\,\mu^{(-2,-2)}\;\Omega\,[{\cal L}^{(2,2)}_{(0)} + 
{\cal L}^{(2,2)}_{(1)} + {\cal L}^{(2,2)}_{(2)} + ... ]~, 
\ee
where ${\cal L}^{(2,2)}_{(0)}$ is just the density in \p{wzwact}.   
Then, using the formula 
$$
\delta\,S_{(0)} = {1\over 4\gamma^2}\;\int \mu^{(-2,-2)}\;\Omega\,
\left( \hat{q}^{(1,1)}\delta\,\hat{q}^{(1,1)}\,{1\over (1+X)^2} \right)~, 
$$
it is rather straightforward to restore the first correction term 
in \p{iterat}: 
\bea
S_{(1)} &=& {1\over 4\gamma^2}\;\int \mu^{(-2,-2)}\;\Omega\,
\left( \hat{q}^{(1,1)}\,{1\over (1+X)^2}\,\left[
c^{(-1,1)}H^{(2,0)}_L + 
c^{(1,-1)}H^{(0,2)}_R \right.\right. \nonumber \\
&& \left.\left. - \;(c^{(1,-1)}H^{(0,2)}_L + 
c^{(-1,1)}H^{(2,0)}_R)(2+X)\right] \right)~.
\eea 
A problem is met at the next step, when trying to calculate the second term. 
Including from the beginning all possible appropriate structures, we 
finally found that almost all structures appearing in the first-order 
variation of $S_{(0)} + S_{(1)}$ can be cancelled by the zeroth-order 
variation of $S_{(2)}$. Only one term cannot be cancelled. It looks 
just the same as the term \p{shift} appearing in 
the variation of the non-invariant $N^{(2,0)}, N^{(0,2)}$ action \p{noninv}
\be  \label{varanom}
\delta\,(S_{(0)} + S_{(1)} + S_{(2)}) = - 
{1\over 4\gamma^2}\;\int \mu^{(-2,-2)}\;\Omega\,\left[ 
2H^{(0,2)}_L\nabla^{(2,0)}\Lambda_L + 2H^{(2,0)}_R\nabla^{(0,2)}\Lambda_R 
\right]~. 
\ee       

The origin of this anomaly can be inferred from the results 
of ref. \cite{hull} where the problem of gauging isometries of bosonic 
sigma models with torsion was studied.  
As was shown there, in the case of group manifold WZW model associated 
with a group $G$ it is impossible to construct an action in which 
the {\it full} $G\times G$ symmetry of the rigid WZW action 
would be gauged (without adding extra copies of WZW fields). 
One can only gauge either the left, or right, or diagonal subgroups 
of $G\times G$. The bosonic sector of the above rigid $q^{(1,1)}$  action 
is just the $SU(2)_L\times SU(2)_R/SU(2)_{diag}$ WZW action, while 
the ``master'' $N=(4,4)$ SG group implies gauging both $SU(2)_L$ and 
$SU(2)_R$ symmetries. Thus, in view of the argument 
just adduced, a direct coupling of the WZW $q^{(1,1)}$ action 
\p{wzwact} to the ``master'' BW multiplet does not exist 
and the ``classical anomaly'' \p{varanom} is just a manifestation 
of this fact. The unremovable piece in the gauge variation \p{shift} 
is of the same origin, because the $N^{(2,0)}, N^{(0,2)}$ action \p{lagrN} 
also contains the $SU(2)_L\times SU(2)_R$ WZW model in its 
bosonic sector. The same reasoning implies the non-existence 
of similar straightforward $N=(4,4)$ SG-II group-invariant 
extensions of \p{wzwact}, \p{lagrN}, since this SG group still 
includes gauge $SU(2)_L, SU(2)_R$ symmetries which act on the 
physical bosons of $q^{(1,1)}$ ($SU(2)$ WZW fields). Note 
that no problems of this sort arise while promoting \p{wzwact} 
to an invariant of the $N=(4,4)$ SG-I gauge group, or to that of the
``master'' SG group with making use of the $N^{(2,0)}, N^{(0,2)}$ 
compensators at the intermediate step: such locally supersymmetric 
$q^{(1,1)}$ actions are particular cases of \p{locqact}, \p{tildeact}.
  
Thus the construction of direct couplings of $N=(4,4)$ WZW action 
\p{wzwact} or the $N^{(2,0)}, N^{(0,2)}$ action \p{lagrN} 
to the ``master'' conformal $N=(4,4)$ SG or $N=(4,4)$ SG-II is 
a non-trivial problem. It seems that the unique possibility to arrange 
such couplings is to consider a few copies of the superfields 
$q^{(1,1)}$, $N^{(2,0)}, N^{(0,2)}$. Then one can construct 
invariant actions as sums of the individual actions of the type 
\p{trialact}, \p{iterat}, taking some of them with the wrong sign 
so as to cancel out the non-vanishing variations like \p{shift}, \p{varanom} 
coming from different actions. This is possible just because these 
anomalous variations involve only SG gauge fields. 

The simplest possibility is to consider a pair of nonlinear multiplets, 
$N_1^{(2,0)}, N_1^{(0,2)}$ and $N_2^{(2,0)}, N_2^{(0,2)}$, each set being 
subjected to the constraints \p{constrNcurv}. 
Then the difference of two actions \p{noninv}
\bea
&& S^{loc}_{N_1N_2} = S^{loc}_{N_1} - S^{loc}_{N_2} = 
-\int \mu^{(-2,-2)}\,\Omega \, \left[ N^{(2,0)}_1 N^{(0,2)}_1 - 
N^{(2,0)}_2 N^{(0,2)}_2 \right. \nn \\ 
&& \left. + \;
(N^{(2,0)}_1 - N^{(2,0)}_2)(H^{(0,2)}_L - H^{(0,2)}_R) - 
(N^{(0,2)}_1 - N^{(0,2)}_2)(H^{(2,0)}_L - H^{(2,0)}_R) \right] \label{N1N2} 
\eea
can be easily checked to be invariant under the ``master'' SG group. 
Each of these multiplets, or their sum 
$N_1^{(2,0)} + N^{(2,0)}_2,  N_1^{(0,2)} + N^{(0,2)}_2$ can be chosen as 
compensators reducing the ``master'' SG group to SG-I group via 
gauge-fixings like \p{gaugeN} (and, further, \p{gaugePhi}). Note 
that the gauge-invariant combinations $
\tilde{N}^{(2,0)} = N^{(2,0)}_1 - N^{(2,0)}_2,  
\tilde{N}^{(0,2)} = N^{(0,2)}_1 - N^{(0,2)}_2$ obey the constraints 
\bea
&& \left[\nabla^{(2,0)} + 
(N^{(2,0)}_1 + N^{(2,0)}_2) \right] \tilde{N}^{(2,0)} = 0\,, \;
\left[\nabla^{(0,2)} + (N^{(0,2)}_1 + N^{(0,2)}_2)\right] 
\tilde{N}^{(0,2)} = 0\,, \nonumber \\
&& \nabla^{(2,0)}\tilde{N}^{(0,2)} - \nabla^{(0,2)}\tilde{N}^{(2,0)} = 0~, 
\label{tildeNconstr} 
\eea
which are recognized as the $\alpha = 0$ version of \p{curGconstr}. 
So one can from the beginning add the invariant piece 
\be
\sim \; \tilde{N}^{(2,0)}\tilde{N}^{(0,2)} \label{N1N2mod}
\ee
to the Lagrangian density in \p{N1N2}. Finally, e.g., in the gauges 
$$
N^{(0,2)}_1 + N^{(0,2)}_2 = N^{(2,0)}_1 + N^{(2,0)}_2 = 0
$$
and \p{gaugePhi}, we are left with the action of the
(32 + 32) matter multiplet 
$\tilde{N}^{(2,0)}, \tilde{N}^{(0,2)}$ in the background of the 
BW-I multiplet augmented with the $U(1)$ multiplet \p{bosfer}. The 
invariant couplings of an arbitrary number of $q^{(1,1)}$ multiplets can 
be arranged with the help of the compensator $N^{(2,0)}_1 + N^{(2,0)}_2, 
N^{(0,2)}_1 + N^{(0,2)}_2 $ as explained above, and added to the 
$N_1, N_2$ action.

Another possibility to set up a direct coupling to the ``master'' $N=(4,4)$ 
conformal SG is to take the difference of the ``almost-covariant'' 
$N$ -superfields action \p{noninv} and the $q^{(1,1)}$ action \p{iterat}
\be
S^{loc}_{qN} = S^{loc}_{wzw} -{1\over 4\gamma^2}\,S^{loc}_N~.   
\label{qN}
\ee 
In analogy with the $N^{(2,0)}, N^{(0,2)}$ action \p{noninv}, it 
is natural to assume that the only unremovable 
term in the ``master'' SG group variation of \p{iterat}     
is given by \p{varanom}, while all higher-order variations 
can be cancelled by inserting into the Lagrangian density the 
appropriate higher-order structures composed out 
of the analytic vielbeins and the superfield $q^{(1,1)}$. 
This still has to be proved (it would be desirable to find out 
the geometric principle behind such a recursion procedure).
\footnote{The $N=(4,4)$, $2D$ WZW - SG couplings were earlier constructed 
using $N=1$ superfields and the conventional $N=4$ superfields in 
\cite{{Kun},{ketov1},{ketov2}}.} 
If such an ``almost-covariant'' $q^{(1,1)}$ action exists, the   
action \p{qN} is invariant like \p{N1N2}, and in the gauges 
\p{gaugeN}, \p{gaugePhi} we arrive at the SG-I group-invariant action of 
the (8+8) multiplet $q^{(1,1)}$ in the background consisting of the BW-I 
gauge multiplet and the extra multiplet \p{bosfer}. Adding other
$q^{(1,1)}$ superfields in a way covariant under ``master''  SG
group can be accomplished, like in the previous case, with making use of the
compensators  $N^{(2,0)}, N^{(0,2)}$. 

An essentially new situation comes out if one directly couples $q^{(1,1)}$
superfields to BW multiplet, without using $N$-superfields. For this purpose
one should take at least two different WZW $q^{(1,1)}$ 
superfields with the same transformation law \p{tranhat} (although with
different sets of constants $c^{ia}$, generally speaking). Under the
assumptions  that the ``almost-covariant'' $q^{(1,1)}$ action $S^{loc}_{wzw}$
exists to all orders in SG fields and that its non-invariance is given only by
the variation \p{varanom}, the fully invariant action could be constructed as  
\be
S^{loc}_{q_1q_2} = S^{loc}_{wzw_1} - S^{loc}_{wzw_2}~. \label{locwzw2}
\ee 
More generally, one can take a sum of $n$ such actions and choose
the coefficients in such a way that the anomaly variations
\p{varanom}  coming from different items in the sum are cancelled out.
\footnote{A similar trick was used in \cite{sazd} for construction of a
gauge-invariant WZW action with the full $G\times G$ symmetry group gauged.}
Clearly, at least one  of such actions should enter with a ``wrong'' sign,
presumably  indicating that the relevant $q^{(1,1)}$ is a sort of  
``Liouville coordinate'' \cite{Kun}. In view of the inhomogeneous nature 
of the transformation law \p{tranhat}, one of 
the $q^{(1,1)}$ superfields will play the role of a compensator.
 
As it was already mentioned, for the time being we are not aware  
of the full nonlinear structure of the ``almost-covariant'' $q^{(1,1)}$ 
actions and even of the complete proof of their existence. Nevertheless, 
taking for granted that such actions can be constructed, let us inspect 
which kind of compensation of the ``master'' SG group can be achieved 
with the help of $q^{(1,1)}$. It will be enough to perform this 
analysis at the linearized level. 

We shall start from the linearized WZ gauge content of 
BW gauge multiplet \p{h20} and the corresponding form \p{l20} of 
the residual symmetry. At the linearized level, the $q^{(1,1)}$ 
constraints \p{curqucons} read (for the shifted 
superfield $\hat{q}^{(1,1)} = q^{(1,1)} - c^{(1,1)}$)
\bea
&& D^{(2,0)}\hat{q}^{(1,1)} = c^{(1,-1)}\, D^{(0,2)}H^{(2,0)}_R - 
c^{(1,1)}\,H^{(2,0)}_R~, \nn \\  
&& D^{(0,2)}\hat{q}^{(1,1)} = c^{(-1,1)}\, D^{(2,0)}H^{(0,2)}_L - 
c^{(1,1)}\,H^{(0,2)}_L~. \label{linconstr} 
\eea
They imply 
\bea
\hat{q}^{(1,1)} &=& \hat{q}^{ia}(z)u^{(1,0)}_iv^{(0,1)}_a + 
\theta^{(1,0)\,\underline{i}}\,\psi^a_{+\,\underline{i}}(z)\,v^{(0,1)}_a 
+
\theta^{(0,1)\,\underline{a}}\,\chi^i_{-\,\underline{a}}(z)\,u^{(1,0)}_i \nn \\
&& + \;i \theta^{(1,0)\,\underline{i}}\,\theta^{(0,1)\,\underline{a}} \, 
F_{\underline{i}\underline{a}}(z) +\; \ldots~, \label{contII}
\eea 
where dots stand for the terms involving the BW multiplet gauge fields 
and derivatives of the explicitly written physical dimension 
fields of $q^{(1,1)}$. The purely shift part of the transformation 
\p{tranhat} (we need only the latter for our linearized analysis) 
\be
\delta\, \hat{q}^{(1,1)} = c^{(1,1)}\,\left(\Lambda_L + \Lambda_R \right) - 
c^{(-1,1)}\,D^{(2,0)}\Lambda_L - c^{(1,-1)}\,D^{(0,2)}\Lambda_R
\label{lintran}
\ee
amounts to the following transformations of the fields:
\bea
&& \delta\,\hat{q}^{ia} = c^{ia}\,(\lambda_L + \lambda_R) - 
c^{\;a}_{j}\,\lambda^{(ji)}_L  -  c^{i}_{\;b}\,\lambda^{(ba)}_R~, \nn \\
&& \delta\, \psi^a_{+\,\underline{i}} = 
- c^{\;a}_{i}\,\beta^i_{+\,\underline{i}}~, 
\quad  
\delta\,\chi^i_{-\,\underline{a}} =  
- c^{i}_{\;a}\,\beta^a_{-\,\underline{a}}~, \nn \\
&& \delta \,F_{\underline{i}\underline{a}} = 0~. \label{tranIIlin}
\eea
One observes that all the physical dimension fields can be gauged away by
appropriate gauge parameters \bea
&&\hat{q}^{ia} = 0 \; \Rightarrow \; (a)\; \lambda_L= -\lambda_R \equiv 
\lambda~, \;\;(b)\; \lambda_L^{(ij)} = {1\over c^2}\,\lambda_R^{(ab)}\,
c^{i}_{\;a} c^{j}_{\;b}  \equiv \lambda^{(ij)}\,, \label{qugauge} \\
&& \psi^a_{\underline{i}} = \chi^i_{\underline{a}} = 0 \quad 
\Rightarrow \quad \beta^i_{+\,\underline{i}} = 
\beta^a_{-\,\underline{a}} = 0~,  \label{fermgauge} 
\eea
where $c^2 = c^{ia}c_{ia} \neq 0$. As it follows from \p{qugauge},  
the product of two local $U(1)$ symmetries is 
compensated down to the diagonal $U(1)$, and the same occurs for 
the product $SU(2)_L\times SU(2)_R$ (this results in   
identifying the $SU(2)$ indices of the left and right harmonics, 
though still does not reduce two harmonic sets to each other). 
Eq. \p{fermgauge} implies the full compensation of the local 
non-canonical supersymmetries. 

As the result, in the gauge \p{qugauge}, \p{fermgauge} the irreducible 
off-shell gauge representation  comprises the (0+0) BW-I multiplet
\p{WZ1SG} as a submultiplet of the ``master'' BW multiplet we started with, as
well as a new 
off-shell (8+8) gauge multiplet. The latter inherits a part of its fields
from  the original BW multiplet, and a part from the compensating 
$\hat{q}^{(1,1)}$ multiplet 
\bea
&&\underline{\mbox{bosons}}: \qquad (h_{++},\; h_{--})\;\; (1,1)~, \quad
(h^{(ik)}_{++},\; h^{(ik)}_{--})\;\;(1,3)~, \quad 
F_{\underline{i}\underline{a}}\;\;(1,4)~,\nonumber \\
&&\underline{\mbox{fermions}}: \qquad 
t^{i\,\underline{a}}_{++-}\;\;(3/2, 4)~, \quad t^{k\underline{j}}_{--+}\;\;
(3/2,4)~. \label{vectmult}
\eea
Comparing it with the (8+8) ``$Sp(1)$ vector multiplet'' of 
ref. \cite{schout}, we find almost full identity between the
two representations, except for a minor distinction related to the fact 
that one bosonic degree of freedom in \p{vectmult} is represented by 
the dimension 1 $U(1)$ gauge field $h_{\pm\pm}$, while in \cite{schout} 
it is carried over by the dimension 2 auxiliary field. It is natural 
to identify the latter with the curl $\partial_{++}h_{--} - 
\partial_{--}h_{++}$, in view of the well-known equivalence of the
auxiliary scalar field and the curl of gauge vector field in two dimensions. 
Note that the $Sp(1)$ vector multiplet was introduced in \cite{schout} 
``by hand'', in addition to the purely 
gauge SG multiplet which we call here BW-I, in order to be able 
to construct locally $N=(4,4)$ supersymmetric sigma models 
on quaternionic manifolds. In our scheme it naturally appears, along 
with the BW-I gauge multiplet, as a result of compensating 
the ``master'' $N=(4,4)$ SG group by the TM-II multiplet $q^{(1,1)}$. 
The gauge $Sp(1)$ symmetry of \cite{schout} is recognized as the  
diagonal in 
the product of $SU(2)_L$ and $SU(2)_R$ symmetries realized as isometries of
the  WZW bosonic fields in $q^{(1,1)}$. It would be of interest to study 
this correspondence at the full nonlinear level and, in particular, 
to inquire how to construct superconformally-invariant couplings of 
some other matter $q^{(1,1)}$ superfields to this field representation 
(different from a simple sum of the ``almost-covariant'' actions). 
Because of the presence of the $SU(2)_{diag}$ gauge fields in \p{vectmult} 
which couple to the physical bosonic fields of $q^{(1,1)}$, 
such couplings should be very restrictive. 

Let us summarize the above ways of descending from the 
``master'' conformal BW multiplet to the BW-I multiplet. 

\vspace{0.4cm}
\noindent{\it A. The (32 + 32) field representation.} This option corresponds 
to the use of the pure gauge nonlinear multiplet 
$N^{(2,0)}$, $N^{(0,2)}$ as the conformal compensator. 
One imposes the covariant constraints \p{constrNcurv} which 
imply some specific form for the analytic vielbeins $H^{(4,0)}$, 
$H^{(0,4)}, H^{(2,2)}, \tilde{H}^{(2,2)}$. After properly fixing the gauges, 
one ends up with the BW-I multiplet and an additional (32+32) off-shell 
$U(1)$ gauge multiplet \p{bosfer} represented by the analytic 
superfield strength $H^{(2,2)}$ \p{defHnew}. The general action 
of the TM-II superfields $q^{(1,1)}$ in the background 
of this representation is given by eq. 
\p{tildeact}. No action for the compensator $N^{(2,0)}$, $N^{(0,2)}$ 
itself is assumed. A version with the additional constraints 
\p{modconstr} yields the pure BW-I multiplet, with no extra multiplets.

\vspace{0.4cm}
\noindent{\it B. The (64+64) field representation.} This case corresponds to 
assuming an invariant action for the  $N^{(2,0)}$, $N^{(0,2)}$ compensator. 
It is constructed using two copies of such superfields 
(eqs. \p{N1N2}, \p{N1N2mod}). Only one set from this pair is the genuine 
compensator. As in the previous case, after gauging this compensator 
away, one ends up with the BW-I multiplet and 
the (32+32) multiplet \p{bosfer}. 
One more (32+32) off-shell multiplet is $\tilde{N}^{(2,0)}$, 
$\tilde{N}^{(0,2)}$  \p{tildeNconstr},which is the remnant of the two
original copies of $N$-multiplets.

\vspace{0.4cm}
\noindent{\it C. The (40+40) field representation.} In this scheme, in
order to construct the invariant action for the compensator  
$N^{(2,0)}$, $N^{(0,2)}$, one uses the hypothetical ``almost-covariant'' 
action for one TM-II multiplet $q^{(1,1)}$ which is 
a gauged extension of the $N=(4,4)$, $SU(2)$ WZW action \p{wzwact}. 
The invariant action of two multiplets 
is given by \p{qN}. After gauging away the
$N$-compensator, one is left with the
(0+0) BW-I multiplet, the (32+32) $U(1)$ multiplet \p{bosfer} and 
the (8+8) TM-II multiplet $q^{(1,1)}$. 

\vspace{0.4cm}
\noindent{\it D. The (16+16) field representation.} This option is different 
 from the preceding ones, as it uses $q^{(1,1)}$ as a compensator 
for the SG-I group. The invariant action \p{locwzw2} is given by the difference
of two  ``almost-covariant'' $q^{(1,1)}$ actions. In the gauge with all
possible  symmetries of the ``master'' SG group being compensated for, the
surviving  field representation consists of the BW-I multiplet, the
(8+8) $SU(2)$ gauge 
multiplet \p{vectmult} and the extra (8+8) TM-II multiplet $q^{(1,1)}$ 
which was added to set up the action \p{locwzw2}.  

\vspace{0.4cm}
There still remain the questions as to, how to descend to another, 
smaller conformal $N=(4,4)$, $SU(2)$ SG group, i.e. the SG-II group,  
and how to reproduce the known \cite{Gates1}-\cite{ketov2} and, perhaps, 
the new $N=(4,4)$ Poincar\'e SG multiplets, by continuing the above 
process of compensation. 

The answer to the first question is as follows. As was already 
mentioned, there should be a ``democracy'' between different 
$SU(2)$ factors in the automorphism group $SO(4)_L\times SO(4)_R$ of 
the $N=(4,4)$, $2D$ Poincar\'e superalgebra. This implies the existence of 
``mirror'' counterparts of the superconformal matter multiplets  
discussed so far, such that the roles of the
$SU(2)$ groups acting on the 
doublet indices $i, a$ and $\underline{i}, \underline{a}$ are 
switched. An example of such a correspondence is the TM-I multiplet 
\cite{{GHR},{IK1},{IKLev1}}, the off-shell field content of which is
given by $q^{\underline{i} \underline{a}}, \psi^{\underline{i}}_k, 
\chi^{\underline{a}}_b, F_{ia}$ that should be compared with 
the field content of TM-II \p{contII}. A similar mirror counterpart should 
exist for the nonlinear multiplet $N^{(2,0)}, N^{(0,2)}$. It is natural 
to call the latter NM-II, with respect to the $N=(4,4),\;SU(2)$ SG-II group
acting on the harmonic variables. Then, with respect to the same SG group,
the mirror counterpart can be called NM-I.  It seems plausible to conjecture
that these mirror TM-I and NM-I  multiplets (being, in fact, TM-II and NM-II
with respect to the $N=(4,4), \, SU(2)$ SG-I group), can be employed to
compensate the ``master'' conformal  SG group just down to the SG-II group,
quite analogously  to how TM-II and NM-II can be used for compensating the
``master''  group down to the SG-I group. Some of these  mirror matter
multiplets, in the rigid case, admit a  description in the $SU(2)\times SU(2)$
harmonic superspace \cite{IS2},  so we can hope to find their locally
supersymmetric versions,  cousins of the actions considered above.  

To clarify the second question, let us come back to the action \p{locwzw2} 
and assume that the ``master'' conformal SG group is reduced 
in it ``by hand'' to the SG-II one (taking for granted that a nonlinear 
version of the truncation conditions \p{trun2H} exists).  One of the
$q^{(1,1)}$ superfields can still be used as a compensator. The linearized, 
purely shift part of the transformation laws of its components, under 
the action of the residual group \p{l2SG}, can be obtained by 
the substitution of \p{trun2par} into \p{tranIIlin} 
\bea
&& \delta\,\hat{q}^{ia} = -{1\over 2}\,c^{ia}\,
(\partial_{++}\lambda^{++} + \partial_{--}\lambda^{--}) - 
c^{\;a}_{j}\,\lambda^{(ji)}_L  -  c^{i}_{\;b}\,\lambda^{(ba)}_R~, \nn \\
&& \delta\, \psi^a_{+\,\underline{i}} = 
2i\, c^{\;a}_{i}\,\partial_{++}\lambda^{+\,i}_{\underline{i}}~, 
\quad  
\delta\,\chi^i_{-\,\underline{a}} =  
2i\,c^{i}_{\;b}\,\partial_{--}\lambda^{-\,b}_{\underline{a}}~, \nn \\
&& \delta \,F_{\underline{i}\underline{a}} = 0~. \label{tranSUlin}
\eea
One sees that the two chiral $SU(2)\,$s are reduced to the diagonal $SU(2)$,
like in the case \p{tranIIlin}, \p{qugauge}, \p{fermgauge},
by gauging away the triplet part of 
$\hat{q}^{ia}$. However, the singlet part cannot be gauged away; it 
becomes just the third component of zweibein. 
Analogously, $\psi^a_{+\,\underline{i}}$, $\chi^i_{-\,\underline{a}}$, together with 
$h^{+\;i\underline{i}}_{--}, h^{-\;a\underline{a}}_{--}$ from the 
BW-II multiplet \p{h2SG}, are combined into the 16-component  
$N=(4,4)$ Poincar\'e SG gravitino (the indices $i$ and $a$ now refer to the 
same diagonal $SU(2)$). Eventually, bearing in mind the auxiliary field 
$F_{\underline{i}\underline{a}}$, we end up just with the $(8+8)$ 
off-shell content of the minimal $N=(4,4)$, $2D$ Poincar\'e SG 
representation \cite{{Gates1},{Gates2}}. However, recalling that 
the invariant action \p{locwzw2} includes one more $q^{(1,1)}$, 
the total off-shell representation for this case is (16+16). This off-shell 
content coincides with that of the ``TM $N=4$ superstring'' 
considered in \cite{Gates3}. 

Analogously, one can use the nonlinear multiplet NM-II as a compensator  
from $N=(4,4)$, $SU(2)$ SG-II down to some Poincar\'e SG. The resulting 
version involves (32+32) off-shell components; its interesting feature 
is that {\it both} conformal $SU(2)$ symmetries turn out to be fully
compensated for, and 
$h^{(ik)}_{++}, h^{(ab)}_{--}$ in \p{h2SG} (and in its left counterpart) 
cease to be gauge fields. The full off-shell content, taking into account 
an additional $q^{(1,1)}$ multiplet needed to construct the
invariant 
action as in \p{qN}, is (40+40). This  coincides with the off-shell
content of  the ``relaxed hypermultiplet $N=4$ superstring'' of ref.
\cite{Gates3}.  It is interesting to inquire whether the latter representation
is indeed  identical to ours. 

At last, one can start from the action \p{N1N2} and recover a version of 
Poincar\'e SG with (64+64) off-shell fields. Once again, in this version both
$SU(2)$  symmetries are fully compensated for.   

In accord with the previous discussion, various mirror versions 
of the Poincar\'e SG can be obtained, starting from the $N=(4,4)$, 
$SU(2)$ SG-I and making use of the multiplets TM-I and NM-I 
as compensators. The various patterns of descent from 
the ``master'' $N=(4,4)$ SG to the SG-I described above, as well as their 
hypothetical mirror cousins, seem also to admit further compensations
down to the $N=(4,4)$ Poincar\'e SG representations, along similar lines.  
New possibilities can arise while simultaneously using both
types of  matter multiplets, i.e. the types I and II, as compensators. 

Finally, let us note that there exists 
a dual version of the rigidly supersymmetric $q^{(1,1)}$  actions, 
including the  $N=(4,4)$ WZW one \p{wzwact}, in terms of unconstrained 
$SU(2)\times SU(2)$ harmonic analytic superfields with infinite numbers of 
auxiliary fields \cite{ISu}. This should obviously generalize to the 
case of local SUSY, which in turn suggests the existence of 
new versions of Poincar\'e $N=(4,4)$, $2D$ SG with infinite sets 
of auxiliary fields.   

A thorough analysis of all these possibilities can be a good 
program for a future study. 

\setcounter{equation}{0}
\section{Conclusions}
In this paper we constructed a new sort of $N=(4,4)$, $2D$ conformal SG 
gauge multiplet, i.e. the Beltrami-Weyl multiplet, starting from the group of 
diffeomorphisms in the $SU(2)\times SU(2)$ analytic harmonic 
superspace. This multiplet can be regarded as the result of gauging
the most 
extensive rigid $N=(4,4)$ superconformal $2D$ group, i.e. the product of two
light-cone copies of the infinite-dimensional 
``large'' $SO(4)\times U(1)\;$, $N=4$ superconformal group. The
previously known $N=(4,4)$  conformal SG groups and the corresponding Weyl
multiplets were argued  to follow from the new ``master'' SG group and BW
multiplet upon their various truncations and compensations, with making use 
of the appropriate superconformal matter multiplets. Also, various 
versions of $N=(4,4)$ Poincar\'e SG can be recovered.

There still remain a few important conceptual and technical points 
to be fully elaborated on. This concerns, before all, constructing the 
full nonlinear version of the ``almost-covariant'' $q^{(1,1)}$ 
action \p{iterat} and the nonlinear completion of the constraints 
\p{trun2H}, \p{trun2}, as well as revealing the component 
fields structure of the locally supersymmetric superfield actions 
presented. An important problem is to incorporate into the present scheme 
mirror counterparts of the superconformal multiplets employed in 
this paper and to study the relevant compensation patterns. 
Different $N=(4,4)$ SG-matter couplings  correspond 
to various versions of $N=(4,4)$ superstrings \cite{Gates3}.  
It would be interesting to inquire the quantum properties of 
the systems described here, e.g., along the lines 
of refs. \cite{IKLev2}, \cite{Kun}. Note that the rigid $N=(4,4)$ WZW 
action \p{wzwact} admits an extension to the
$N=(4,4)$ WZW-Liouville one \cite{{IK1},{IKLev1},{GorI},{ISu}}, 
with breaking the $N=(4,4),\; SO(4)\times U(1)$ 
superconformal invariance down to the type-II $N=(4,4),\; SU(2)$ 
one. Such a Liouville extension plays an important role in the 
quantum case \cite{Kun}. It is of interest to inquire whether 
a locally supersymmetric extension of the Liouville term can be 
constructed in $SU(2)\times SU(2)$ harmonic superspace.

\section*{Acknowledgments}
S.B. wishes to thank JINR-Dubna for hospitality at the early stages
of this research.
E.I. thanks Jim Gates for enlightening correspondence and INFN-LNF
for the hospitality multiply extended to him during
the course of this long-term work. This research was supported in part
by the Fondo Affari Internazionali Convenzione 
Particellare INFN-JINR, Project PAST-RI 99/01, RFBR Grant 99-02-18417, 
RFBR-CNRS Grant 98-02-22034, NATO Grant PST.CLG 974874 and INTAS 
Grants INTAS-96-0538, INTAS-96-0308. 

\setcounter{equation}0
\def\theequation{A.\arabic{equation}}
\section*{Appendix: A simple example of the component action}
Here, just to give a feeling how the locally supersymmetric 
actions presented in this paper look in terms of component fields, we 
quote the free part of the general conformal 
SG-I group-invariant action \p{locqact}
\be
S_q^{free} = - \int \mu^{(-2,-2)}\;
\hat{\Omega}\;q^{(1,1)}\,q^{(1,1)}~. \label{qfree}
\ee

It will be convenient to choose a gauge for the analytic vielbeins 
which is slightly different from \p{gauges1SG}
\bea
&& H^{(2,0)\,++} = H^{(3,0)\,\underline{i}} = 0~, \quad 
H^{(2,0)\,--} = i(\theta^{(1,0)})^2\,\hat{h}^{--}_{++}(z,v,\theta^{(0,1)})~, 
\nn \\ 
&& H^{(2,1)\,\underline{a}} = 
i(\theta^{(1,0)})^2\,\hat{h}^{(0,1)\,\underline{a}}_{++}(z,v,\theta^{(0,1)})~.
\eea
This gauge is also globally well-defined. To simplify the situation as
soon as possible, we recall that all components of the BW-I multiplet 
are locally pure gauge, and we choose the additional gauge, 
which is admissible only locally,
\be
h^{--}_{++} = h_{++}^{(\underline{a}\underline{b})} = 
h^{-\,a\underline{a}}_{++} = 0 \label{rightG}
\ee
(we could alternatively choose the left counterparts 
of \p{rightG} to vanish). It is easy to show that the 
{\it full} solution of the constraints \p{constr1SG} in this gauge 
is given by 
\bea
\nabla^{(2,0)} &=& \partial^{(2,0)} + i(\theta^{(1,0)})^2 \,\partial_{++}~, 
\;\;
\nabla^{(0,2)} = \partial^{(0,2)} + 
i(\theta^{(0,1)})^2 \,\nabla_{--}~, \nn \\
\nabla_{--} &=& \partial_{--} + \{\,h^{++}_{--} - 2i\,
\theta^{(1,0)}_{\underline{i}}\,h^{+\;k\underline{i}}_{--}u^{(-1,0)}_k\,\} 
\,\partial_{++}  \nn \\
&+&  \{\, h^{+\;k\underline{i}}_{--}u^{(1,0)}_k + 
\theta^{(1,0)\;\underline{k}} 
[ h^{(\underline{i}}_{--\;\underline{k})} +{1\over 2}\delta^
{\underline{i}}_{\underline{k}}\;\partial_{++} h^{++}_{--}] \nn \\
&-&  (\theta^{(1,0)})^2 \partial_{++}h^{+\;k\underline{i}}_{--} 
v^{(-1,0)}_k\, \} \frac{\partial}{\partial \theta^{(1,0)\;\underline{i}}}~. 
\label{nablyG}
\eea
It is easy to explicitly check the integrability condition 
$$
[\,\nabla^{(2,0)}, \nabla^{(0,2)} \,] = 0~.
$$
The residual gauge symmetry of \p{nablyG} is given by \p{Res1SG},
with all parameters being functions of only $z^{--}$ (this is just the
right $N=4,\;SU(2)$ SCA-I), and by the left 
counterpart of \p{Res1SG}, with the parameters still being general 
functions of both coordinates $z^{\pm\pm}$. It is easy to check that under this
group  $$
\delta \mu^{(-2,-2)} = 0~, 
$$
so one can expect $\hat{\Omega} \sim \mbox{const}$ in this gauge. 
This is indeed so, because it is easy to check that  
\be
\Gamma^{(2,0)} = \Gamma^{(0,2)} = 0
\ee
for the vielbeins in \p{nablyG}. Then, the action  \p{qfree} is 
\be
S_q^{free} = - \int \mu^{(-2,-2)}\;q^{(1,1)}\,q^{(1,1)}~, \label{qfree1}
\ee
with
\be
\nabla^{(2,0)}q^{(1,1)} = \nabla^{(0,2)}q^{(1,1)} = 0~. \label{constrG}
\ee
Being aware of the explicit expressions for $\nabla^{(2,0)}, \nabla^{(0,2)}$, 
it is easy to directly solve these constraints in terms of the physical 
fields of $q^{(1,1)}$ defined in \p{contII} and the SG fields. This is
rather straightforward, so we quote only the final form of the action. 
It is obtained by substituting this solution into \p{qfree1} and integrating
there  over the $\theta$'s and the harmonics:  
\bea
S_q^{free} &=& \int d^2 z 
\left\{ \partial_{++}q_{ia}\,(\hat{\nabla}_{--}q^{ia} 
+ h_{--}^{+\;i\underline{i}}\,\psi^{a}_{+\,\underline{i}}) + 
{i\over 2}\, \chi^{i\underline{a}}_-\partial_{++} \chi_{-\,i\underline{a}} 
+ {1\over 4}\, 
F^{\underline{i}\underline{a}}F_{\underline{i}\underline{a}}  
\right. \nn \\
&+& \left.{i\over 2}\,\psi^{a\underline{i}}_+\left[ 
\hat{\nabla}_{--} \psi_{+\;a\underline{i}} + 
(h^{(\underline{k}}_{--\,\underline{i})} + 
{1\over 2}\delta^{\underline{k}}_{\underline{i}}\, 
\partial_{++}h^{++}_{--})\,\psi_{+\,a\underline{k}} -2i 
h^{+\,k}_{--\,\underline{i}}\partial_{++} q_{ka} \right] \right\},
\label{chirN4}
\eea
where 
$$
\hat{\nabla}_{--} = \partial_{--} + h^{++}_{--}\partial_{++}~.
$$ 
Note that \p{chirN4} is just the action of the
$N=4$ chiral bosons constructed 
in ref. \cite{lab} (up to switching the $+$ and $-$ light-cone 
indices), with the residual local $N=4$ SUSY as the 
relevant Siegel symmetry.

\end{document}